\documentclass[fleqn,usenatbib]{mnras}

\usepackage{graphicx}	
\usepackage{amsmath}	
\usepackage[dvipsnames]{xcolor} 
\usepackage{amssymb}	
\usepackage{newtxtext,newtxmath}
\usepackage{slashbox,multirow,booktabs} 
\usepackage{enumitem} 
\usepackage[flushleft]{threeparttable} 
\usepackage{ulem}

\newcommand{\new}[1]{{\color{Plum}\textbf{#1}}}

\newcommand{\SatGen}{{\tt SatGen}\,}

\newcommand{\eq}[1]{equation~(\ref{eq:#1})}
\newcommand{\eqs}[1]{equations~(\ref{eq:#1})}
\newcommand{\Eqs}[1]{Equations~(\ref{eq:#1})}
\newcommand{\se}[1]{Section \ref{sec:#1}}
\newcommand{\app}[1]{Appendix \ref{app:#1}}
\newcommand{\fig}[1]{Fig.~\ref{fig:#1}}
\newcommand{\figs}[1]{Figs.~\ref{fig:#1}}
\newcommand{\tab}[1]{Table~\ref{tab:#1}}
\newcommand{\be}{\begin{equation}}
\newcommand{\ee}{\end{equation}}
\newcommand{\bad}{\begin{equation} \begin{aligned}}
\newcommand{\ead}{\end{aligned} \end{equation}}

\newcommand{\Msun}{M_\odot}

\newcommand{\kpc}{\,{\rm kpc}}

\newcommand{\rhoc}{\rho_{\rm crit}}
\newcommand{\rhobar}{\overline{\rho}}
\newcommand{\rhog}{\rho_{\rm gas}}

\newcommand{\tdyn}{t_{\rm dyn}}
\newcommand{\deltac}{\delta_{\rm c}}
\newcommand{\Mv}{M_{\rm vir}}
\newcommand{\Ms}{M_{\star}}

\newcommand{\Mc}{M_{\rm c}}
\newcommand{\Md}{M_{\rm d}}
\newcommand{\mv}{m_{\rm vir}}
\newcommand{\ma}{m_{\rm acc}}

\newcommand{\mmax}{m_{\rm max}}
\newcommand{\ms}{m_\star}
\newcommand{\mg}{m_{\rm gas}}
\newcommand{\fg}{f_{\rm gas}}
\newcommand{\fb}{f_{\rm bar}}
\newcommand{\rv}{r_{\rm vir}}
\newcommand{\reff}{r_{\rm eff}}
\newcommand{\rs}{r_{\rm s}}
\newcommand{\rmax}{r_{\rm max}}
\newcommand{\rmin}{r_{\rm min}}

\newcommand{\rc}{r_{\rm circ}}
\newcommand{\lv}{l_{\rm vir}}
\newcommand{\lt}{l_{\rm t}}
\newcommand{\lRP}{l_{\rm RP}}
\newcommand{\leff}{l_{\rm eff}}
\newcommand{\lmax}{l_{\rm max}}
\newcommand{\Vc}{V_{\rm circ}}
\newcommand{\vc}{v_{\rm circ}}
\newcommand{\Vv}{V_{\rm vir}}
\newcommand{\Vmax}{V_{\rm max}}
\newcommand{\Vt}{V_{\rm t}}
\newcommand{\VR}{V_R}
\newcommand{\Vz}{V_z}
\newcommand{\vmax}{v_{\rm max}}
\newcommand{\Vphibar}{\overline{V}_\phi}
\newcommand{\Vphi}{V_\phi}
\newcommand{\jc}{j_{\rm circ}}
\newcommand{\aDF}{\boldsymbol{a}_{\rm DF}}
\newcommand{\lnL}{\ln\Lambda}
\newcommand{\erf}{{\rm erf}}
\newcommand{\xs}{x_{\rm s}}

\newcommand{\xp}{x^\prime}
\newcommand{\xc}{x_{\rm circ}}
\newcommand{\za}{z_{\rm acc}}

\newcommand{\rmd}{{\rm d}}

\defcitealias{NFW97}{NFW}
\defcitealias{Dekel17}{Dekel+}
\defcitealias{Einasto65}{Einasto}
\defcitealias{MN75}{MN}

\begin{document}



\title[SatGen]{SatGen: a semi-analytical satellite galaxy generator -- I. The model and its application to Local-Group satellite statistics}

\author[F. Jiang et al.]{%
Fangzhou Jiang,$^{1,2,3}$\thanks{Troesh Scholar; E-mail: \href{mailto:fzjiang@caltech.edu}{fzjiang@caltech.edu}}
Avishai Dekel,$^{3,4}$
Jonathan Freundlich,$^{3,5}$
Frank C. van den Bosch,$^{6,7}$
\newauthor
Sheridan B. Green,$^{7}$\thanks{NSF Graduate Research Fellow}
Philip F. Hopkins,$^{1}$
Andrew Benson,$^{2}$
and
Xiaolong Du$^{2}$
\vspace*{8pt}
\\
$^{1}$ TAPIR, California Institute of Technology, Pasadena, CA 91125, USA\\
$^{2}$ Carnegie Observatories, 813 Santa Barbara Street, Pasadena, CA 91101, USA \\
$^{3}$ Center for Astrophysics and Planetary Science, Racah Institute of Physics, The Hebrew University, Jerusalem 91904, Israel\\
$^{4}$ SCIPP, University of California, Santa Cruz, CA 95064, USA\\
$^{5}$ School of Physics and Astronomy, Tel Aviv University, Tel Aviv 69978, Israel \\
$^{6}$ Department of Astronomy, Yale University, P.O. Box 208101, New Haven, CT 06520-8101\\
$^{7}$ Department of Physics, Yale University, PO. Box 208120, New Haven, CT 06520-8120\\
}

\date{}

\pubyear{2020}

\label{firstpage}
\pagerange{\pageref{firstpage}--\pageref{lastpage}}
\maketitle

\begin{abstract}
We present a semi-analytical model of satellite galaxies, \SatGen, which can generate large statistical samples of satellite populations for a host halo of desired mass, redshift, and assembly history. 
The model combines dark-matter (DM) halo merger trees, empirical relations for the galaxy-halo connection, and analytical prescriptions for tidal effects, dynamical friction, and ram pressure stripping.
\SatGen emulates cosmological zoom-in hydro-simulations in certain aspects. 
Satellites can reside in cored or cuspy DM subhaloes, depending on the halo response to baryonic physics that can be formulated from hydro-simulations and physical modeling. 
The subhalo profile and the stellar mass and size of a satellite evolves depending on its tidal mass loss and initial structure. 
The host galaxy can include a baryonic disc and a stellar bulge, each described by a density profile that allows analytic satellite orbit integration. 
\SatGen complements simulations by propagating the effect of halo response found in simulated field galaxies to satellites (not properly resolved in simulations) and outperforms simulations by sampling the halo-to-halo variance of satellite statistics and overcoming artificial disruption due to insufficient resolution.
As a first application, we use the model to study satellites of Milky Way (MW) and M31 sized hosts, making it emulate simulations of bursty star formation and of smooth star formation, respectively, and to experiment with a disc potential in the host halo.
We find that our model reproduces the observed satellite statistics reasonably well.
Different physical recipes make a difference in satellite abundance and spatial distribution at the 25\% level, not large enough to be distinguished by current observations given the halo-to-halo variance. 
The MW/M31 disc depletes satellites by ${\sim} 20\%$ and has a subtle effect of diversifying the internal structure of satellites, which is important for alleviating certain small-scale problems. 
We discuss the conditions for a massive satellite to survive in MW/M31.
\end{abstract}

\begin{keywords}
galaxies: dwarf --
galaxies: evolution --
galaxies: haloes --
galaxies: interactions --
galaxies: structure --
methods: numerical
\end{keywords}



\section{Introduction}
\label{sec:intro}

In our modern understanding of the Universe, structures form hierarchically: dark matter (DM) overdensities collapse into gravitationally bound haloes, which merge to form larger haloes. 
The smaller participant of a merger survive as substructure within the merger remnant, experiencing tidal interactions, losing mass, and undergoing structural change. 
Galaxies form inside DM haloes.
When a halo merger occurs, the less massive progenitor becomes a substructure and the inhabiting galaxy becomes a satellite galaxy. 
Subhaloes and satellites are therefore the building blocks of host haloes and central galaxies and serve as relics of structures that formed earlier, with their demographics containing the information of the assembly history of the host system as well as the Universe at large.  

Apart from their cosmological significance, satellite galaxies are interesting on their own, in the sense that galaxies of extreme morphology are usually spotted in dense environments.  
For example, among bright dwarfs (i.e., galaxies with stellar mass $\ms\sim10^{7-9}\Msun$) in the Local Group or in galaxy clusters, galaxies range from ultra-compact dwarfs (UCDs, with half-stellar-mass radii $\leff\sim0.1\kpc$, e.g., \citealt{Drinkwater03}) to ultra-diffuse galaxies (UDGs, with $\leff\sim5\kpc$, e.g., \citealt{vD15}), spanning almost 2 dex in size.
The environment may be the key to such diversity: the central galaxy and the host halo can make a satellite more diffuse or more compact through tidal effects depending on the initial conditions, the time since the infall of the satellite, and the orbit of the satellite.

Subhaloes and satellites have been studied using numerical simulations  \citep[e.g.][]{Gao04, Diemand08, Springel08, Wu13, GK14a, Mao15, Sawala15, Wetzel16, GK19} and semi-analytical models \citep[e.g.][]{TB01, Benson02a, Benson02b, ZB03, Zentner05, Gan10, JB16, Nadler19, Yang20}. 
Cosmological $N$-body simulations produce a plethora of subhaloes compared to observed satellite galaxies.
While low-mass haloes ($\Mv\la10^{9}\Msun$) are expected to be truly dark due to the suppression of star formation by the cosmic UV background, thereby alleviating this ``missing satellite'' problem \citep[e.g.][]{Benson02a, Benson02b, Hambrick11}, a more persistent challenge lies in the overabundance of massive and dense subhaloes -- they are too big to fail forming stars \citep{BK11}.
The ``too-big-to-fail'' problem is not merely the overabundance of massive satellites, but also highlights the lack of structural diversity in the simulated satellite populations \citep[e.g.][]{JB15} --  the simulated population of massive satellites are dense in their centres, showing a narrow distribution of maximum circular velocities ($\vmax$), while the observed bright dwarf satellites exhibit a larger variety of inner densities \citep{Oman15} and a broad distribution of $\vmax$.
Hydro-simulations have shown that including baryons can help to reduce the abundance of massive satellites, mostly because the central galaxies enhance the tidal disruption of satellites (e.g., \citealt{GK19}, but see also \citealt{Errani17} and \citealt{GK17}, which use idealized $N$-body simulations with a galactic disc). 
However, hydro-simulations still do not fully reproduce the structural diversity of dwarf satellites \citep[e.g.][]{GK19}, missing the most diffuse and most compact dwarf satellites seen around the Milky Way (MW) and M31. 

The limitations of cosmological simulations can be summarized as follows. 
First, simulating a satellite population is computationally expensive -- it requires a large dynamical range in mass and in spatial scale.
State-of-the-art zoom-in simulations typically produce on the order ${\sim}10$ MW-like host systems \citep[e.g.][]{Sawala15, GK19} or ${\sim}1$ cluster \citep[e.g.][]{Pillepich19, Tremmel19}, whereas quantifying the cosmic variance of satellite statistics for a given host mass requires at least hundreds of random realizations \citep{PZ12,JB15}. 
Second, artificial disruption of satellites due to insufficient resolution is still prevalent in modern simulations. 
It is alarming to realize that, in the Bolshoi simulation \citep{Klypin11}, ${\sim}60\%$ of subhaloes with infall mass larger than 10\% of the instantaneous host halo mass cannot even survive for one orbit \citep{JB17} and ${\sim}13\%$ of subhaloes are disrupted per Gyr \citep{vdB17}, despite the use of a sophisticated, phase-space based halo finder \citep{Behroozi11}.
Similar results have been reported for zoom-in simulations: about half of the subhaloes in the Aquarius simulations have been disrupted, irrespective of their masses at infall \citep{Han16}. 
Idealized simulations (of higher resolution than cosmological ones) reveal that satellite disruption is mostly numerical in origin, caused mainly due to inadequate force softening and a runaway instability triggered by the amplification of discreteness noise in the presence of a tidal field
\citep{BO18, vdB18}.
Third, halo finding algorithms, especially those based only on identifying instantaneous overdensities, have difficulty in recovering subhaloes when they are located in dense region of the host \citep{Muldrew11, BJ16}.  

Semi-analytical models serve as complementary tools to simulations in the study of satellite galaxies and outperform simulations in terms of statistical power and numerical resolution. 
Such models consist of halo merger trees and analytical prescriptions for satellite evolution.
Most of these models focus on the DM components, using cuspy profiles \citep{NFW97} to describe both the host halo and the satellites, ignoring baryonic components and processes.
However, hydro-simulations have shown that baryonic influence cannot be neglected for satellites.  
First, the DM profile of satellites at infall is not necessarily cuspy. 
For example, supernovae-driven gas outflows can create dark matter cores \citep[e.g.,][]{PG12} and systems with cored profiles follow different tidal evolution paths than cuspy ones with the same initial orbit \citep[e.g.,][]{Penarrubia10}.
Second, the central galaxy, e.g., a MW-like disc, can significantly impact the spatial distribution of a satellite population by reducing the survivability of the satellites that travel across the disc-dominated region \citep[e.g.,][]{GK17}. 
Finally and obviously, to study the baryonic properties of satellite galaxies instead of merely the statistics of DM subhaloes, the baryonic components of a satellite and their evolution in a dense environment must be considered. 
Hence, semi-analytical models of satellites are urgently in the need of upgrades in order to catch up with recent developments in cosmological simulations. 

In this paper, we present \SatGen, a new semi-analytical model for generating merger trees and evolving satellite populations, and then, as a proof-of-concept for \SatGen, we perform a study of satellite statistics for MW/M31-like hosts.
Compared to previous models, \SatGen improves on several important aspects. 
First, it considers baryonic effects, both within the satellites and the host galaxy, on the structure and survivability of subhaloes.
Subhaloes in \SatGen can be described by profiles that have the flexibility to capture DM cores and that have been widely used to describe subhaloes in simulations, including a subclass of the $\alpha\beta\gamma$ profiles \citep{Zhao96, Dekel17, Freundlich19,Freundlich20} and the \citet{Einasto65} profile. 
The initial structure of the subhaloes are based upon halo response models extracted from state-of-the-art hydro-simulations and analytical modeling; by changing the halo response model, the user can make \SatGen emulate different simulations.
Host systems in \SatGen can be composed of (a combination of) a baryonic disc, stellar bulge, and DM halo.
Second, \SatGen incorporates simple recipes for the evolution of the stellar and gaseous components of satellite galaxies.  
The structural evolution recipes of subhaloes and stellar components are either analytical and physically motivated or extracted from high-resolution idealized simulations, which makes \SatGen essentially free from the effects of numerical disruption of satellites commonly seen in cosmological simulations. 
Finally, in keeping with the most sophisticated previous models of this kind \citep[e.g.,][]{TB01, Benson02a, Zentner05}, \SatGen follows the orbit of each satellite, while accounting for dynamical friction. 

This paper is organized as follows. 
In \se{model}, we describe the model.
In \se{application}, we present satellite statistics of MW/M31-sized systems, comparing model predictions with observations (\se{SHMF}), as well as comparing model results using different halo response models characteristic of different hydro-simulations (\se{DifferentHaloResponse}).
We also quantify the effect of a baryonic disc potential on the abundance, spatial distribution, and internal structure of satellites (\se{DiscEffect}).
In \se{discussion}, we explore the conditions for a massive satellite to survive (or get disrupted) in a MW/M31 potential.
In \se{conclusion}, we summarize the model and our findings.

Throughout, we use $m$ and $M$ to indicate satellite mass and host mass, respectively.
We use $l$ and $r$ to refer to satellite-centric radius and host-centric distance, respectively. 
Thus, a density profile written as $\rho(r)$ refers to that of the host system and written as $\rho(l)$ refers to that of the satellite. 
We define the virial radius of a distinct halo as the radius within which the average density is $\Delta=200$ times the critical density for closure. 
We adopt a flat $\Lambda$CDM cosmology with the present-day matter density $\Omega_{\rm m}=0.3$, baryonic density $\Omega_{\rm b}=0.0465$, dark energy density $\Omega_\Lambda=0.7$, a power spectrum normalization $\sigma_8=0.8$, a power-law spectral index of $n_s=1$, and a Hubble parameter of $h=0.7$.
All of these assumptions can be changed easily in \SatGen. 

\section{Model}
\label{sec:model}

\begin{figure*}
	\includegraphics[width=\textwidth]{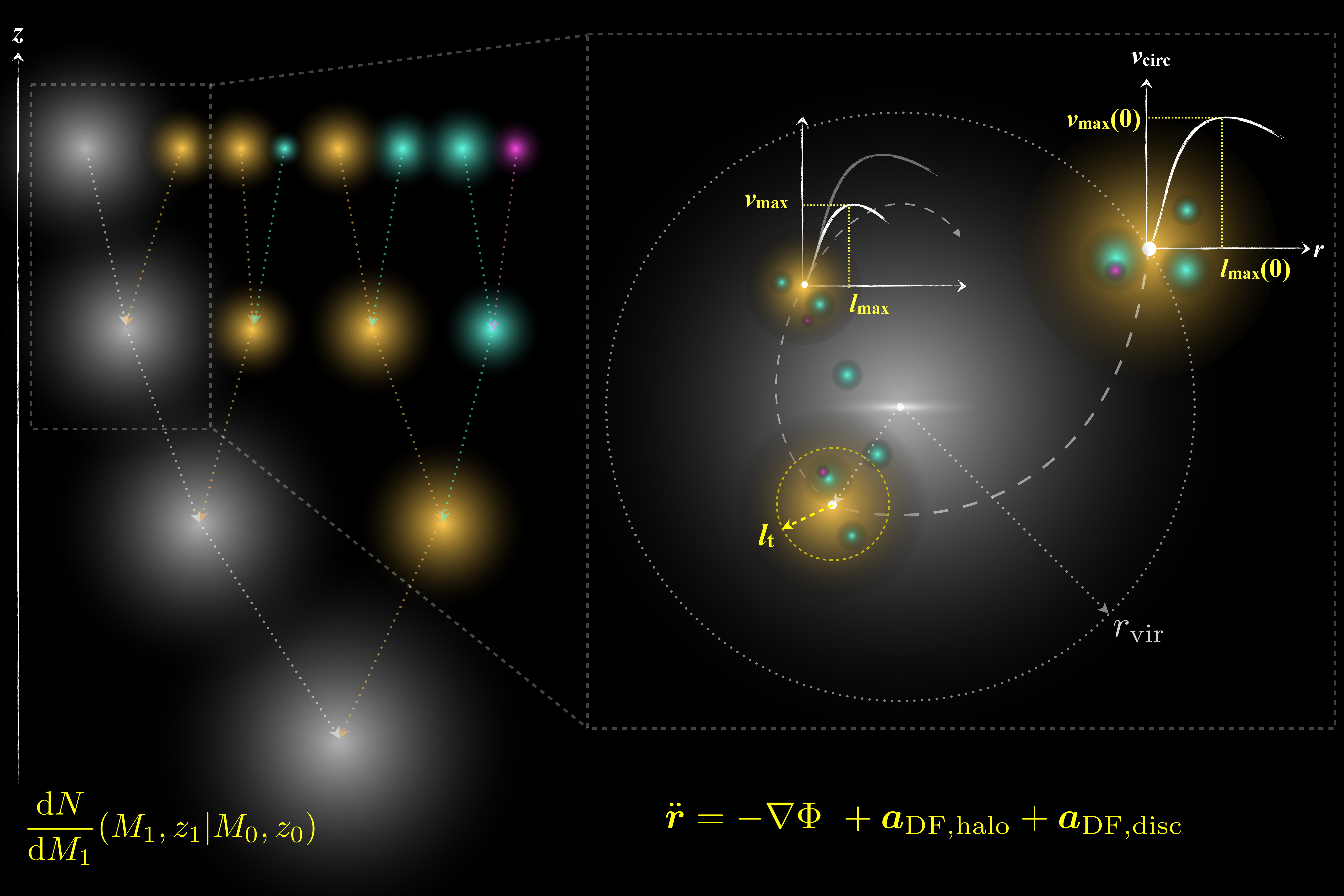}
    \caption{Schematic view of the \SatGen model. {\it Left}: a halo merger tree, generated by sampling the EPS progenitor mass function, $\rmd N / \rmd M_1 (M_1,z_1|M_0,z_0)$ (see \se{tree}). Different colours differentiate branches of different levels -- the main branch (i.e., the host-halo branch) is white; the branches of the first-order satellites, i.e., the satellites that are directly accreted by the host system, are yellow; the branches of the second-order satellites, i.e., the satellites that directly merge with first-order satellite progenitors and are brought into the host halo as sub-substructures, are cyan; and so on. {\it Right}: a zoom-in view of what happens after a satellite is accreted. In this illustration, a first-order satellite orbits around a host composed of a smooth halo and a galactic disc (see \se{initial} for how we initialize the host). The satellite brings its own higher-order substructure to the host, loses mass (see \se{stripping} for how we model tidal stripping), releases higher-order satellites, and evolves in structure (represented by the peak circular velocity, $\vmax$, and the corresponding location, $\lmax$), as illustrated by the schematic plots of the circular velocity profiles at infall (see \se{initial} for how we initialize subhalo structure at infall) and at a later epoch when it is significantly stripped (see \se{tracks} for how the structural evolution is modeled). For such an eccentric orbit (see \se{initial} for how we draw initial orbits), tidal stripping is most efficient at the orbital pericentre, where the Hill surface is indicated by a yellow dotted circle and the tidal radius, $\lt$, is marked (see \se{stripping} for how we model tidal stripping). For such a major merger, orbital decay due to dynamical friction (\se{orbit}) is significant, as illustrated by the dashed line. Not shown here are the prescriptions for the initialization and the evolution of the stellar and gaseous components of the satellite (see \se{initial} and \se{tracks} for details).}
    \label{fig:cartoon}
\end{figure*}

The model builds upon halo merger trees. 
Combining these merger trees with some empirical prescriptions from simulations, we obtain the initial masses, profiles, and baryonic properties of satellites. 
Then, we follow the orbits of the satellites, modeling tidal stripping and the structural evolution of both the DM and baryonic components. 
The \SatGen code is made publicly available on GitHub.\footnote{ \href{https://github.com/shergreen/SatGen}{https://github.com/shergreen/SatGen}}
A schematic view of the model is presented in \fig{cartoon}.
Below, we introduce each model component in sufficient detail to reproduce the exercise in \se{application}, leaving more comprehensive details in the appendices.
Readers who want to see the results first with a basic idea of how the model works can view \fig{cartoon} and read \se{workflow} for a quicker overview and jump to \se{application}.

\subsection{Halo merger trees}
\label{sec:tree} 

\SatGen generates halo merger trees using an algorithm \citep{PCH08} based on the extended Press-Schechter (EPS) formalism \citep{LC93}.
The EPS method provides the expected number of progenitor haloes of mass $M_1$ at redshift $z_1$ for a target halo of mass $M_0$ at redshift $z_0 < z_1$, 
\be
\label{eq:PMF}
\frac{\rmd N}{\rmd M_1}(M_1,z_1|M_0,z_0)\rmd M_1 = \frac{M_0}{M_1} \frac{\Delta\omega}{\sqrt{2\pi}(\Delta S)^{3/2}}e^{-\frac{(\Delta\omega)^2}{2\Delta S}}\left|\frac{\rmd S}{\rmd M}\right |_{M_1}\rmd M_1,
\ee
where $S\equiv\sigma^2(M)$ is the variance of the density field linearly extrapolated to $z=0$ and smoothed with a sharp $k$-space filter of mass $M$, $\delta(z)$ is the critical overdensity for spherical collapse, $\Delta S=\sigma^2(M_1)-\sigma^2(M_0)$, and $\Delta\omega=\deltac(z_1)-\deltac(z_0)$.

However, it has been shown that merger trees constructed by strictly sampling this progenitor mass distribution over-predict the low-redshift merger rate compared to cosmological simulations \citep[e.g.,][]{Zhang08, JB14}.
In order to achieve better agreement with simulations, we follow \citet{PCH08} by adding a corrective factor of the following form to the right-hand side of \eq{PMF}:
\be
G(M_1|M_0,z_0) = G_0 \left(\frac{S_1}{S_0}\right)^{\frac{\gamma_1}{2}}  \left(\frac{\omega_0^2}{S_0}\right)^{\frac{\gamma_2}{2}},
\ee
where $S_1=\sigma^2(M_1)$, $S_0=\sigma^2(M_0)$, $\omega_0=\delta^2(z_0)$, and we adopt $G_0=0.6353$, $\gamma_1=0.1761$, and $\gamma_2=0.0411$ following \citet{Benson17}.

We construct merger trees using the time-stepping advocated in Appendix A of \citet{PCH08}, which corresponds to $\Delta z{\sim}0.001$. 

In order to reduce memory usage, we follow \cite{JB16} and down-sample
the temporal resolution of the trees by only registering progenitor haloes every timestep of $\Delta t = 0.1 \tdyn(z)$, where $\tdyn=\sqrt{3\pi/[16G \Delta \rhoc(z)]}$ is the instantaneous dynamical time of DM haloes.

\subsection{Profiles for DM haloes and baryonic discs}
\label{sec:profiles}

With \SatGen, one has multiple choices for the profile of a DM halo, including the \citet[][hereafter \citetalias{NFW97}]{NFW97} profile, the \citet[][hereafter \citetalias{Einasto65}]{Einasto65} profile, and the \citet[][hereafter \citetalias{Dekel17}]{Dekel17} profile, which is a subclass of the $\alpha\beta\gamma$ profiles \citep{Zhao96}. 
Galactic discs and bulges can be described by the \citet[][hereafter \citetalias{MN75}]{MN75} profile and the \citetalias{Einasto65} profile, respectively.
One can set up a host system using a combination of the aforementioned profiles, e.g., a \citetalias{NFW97} halo plus an embedded \citetalias{MN75} disc.
In \app{profiles}, we provide analytical expressions for the profiles of density, enclosed mass, gravitational potential, and velocity dispersion of all of the supported profiles. 
Here, we describe the \citetalias{Dekel17} halo profile and the \citetalias{MN75} profile, which will be used in the experiments in \se{application}.

\subsubsection{\citet{Dekel17} halo profile}

A \citetalias{Dekel17} halo is defined by four parameters: the virial mass, $\Mv$, a concentration parameter, $c$, the (negative of the) logarithmic density slope in the centre, $\alpha=-\rmd \ln\rho/\rmd \ln r|_{r\to 0}$, and the spherical overdensity, $\Delta$.
The density profile is given by:
\be
\label{eq:DekelDensity}
\rho(r) = \frac{\rho_0}{x^\alpha(1+x^{1/\beta})^{\beta(\gamma-\alpha)}}, \quad \beta=2, \quad  \gamma=3+\beta^{-1}=3.5,
\ee
where $x\equiv r/\rs$ is the radius scaled by an intermediate radius $\rs$ that is related to $\rv$ by the concentration parameter, $\rs = \rv/c$, and $\rho_0 = [c^3(3-\alpha)/3/f(c,\alpha)]\Delta\rhoc$, with $f(x,\alpha)=\chi^{2(3-\alpha)}$ and $\chi\equiv x^{1/2}/(1+x^{1/2})$.

The \citetalias{Dekel17} profile has only one more degree of freedom than the \citetalias{NFW97} profile and it has three merits that make it ideal for use in semi-analytical models. 
First, it can accurately describe haloes in hydro-simulations \citep{Dekel17,Freundlich20}, having enough flexibility near the centre to accurately describe the cusp-core transformation \citep{Freundlich19}. 
Second, it has an outer slope of $\gamma=3.5$, steeper than that of the \citetalias{NFW97} profile and thus more appropriate for describing subhaloes that are stripped.
Finally, it has fully analytical expressions for the profiles of enclosed mass, gravitational potential, and velocity dispersion, facilitating fast orbit integration and making it more convenient to use than the \citetalias{Einasto65} profile or other subclasses of the $\alpha\beta\gamma$ family \citep[see more details in][]{Freundlich20}. 
\footnote{In fact, a full family of profiles of the form of \eq{DekelDensity} with $\beta=n$ and $\gamma = 3+k/n$ (where $k$ and $n$ are integers) have fully analytical expressions for the profiles of potential and velocity dispersion \citep{Zhao96}. The choice of $n=2$ and $k=1$, as in the \citetalias{Dekel17} profile, yields accurate enough descriptions of haloes in hydro-simulations.}
The mass inside radius $r$ is given by
\be
\label{DekelMass}
M(r) = \Mv \frac{f(x,\alpha)}{f(c,\alpha)},
\ee
the gravitational potential can be expressed as 
\be
\label{DekelPotential}
\Phi(r) = -\Vv^2\frac{2c}{f(c,\alpha)}\left[\frac{1-\chi^{2(2-\alpha)}}{2(2-\alpha)}-\frac{1-\chi^{2(2-\alpha)+1}}{2(2-\alpha)+1}\right],
\ee
where $\Vv$ is the virial velocity,
and the one-dimensional isotropic velocity dispersion $\sigma(r)$ is given by
\be
\label{eq:DekelVD}
\sigma^2(r) = 2\Vv^2 \frac{c}{f(c,\alpha)} \frac{x^{3.5}}{\chi^{2(3.5-\alpha)}} \sum_{i=0}^{8} \frac{(-1)^i 8!}{i!(8-i)!}  \frac{1-\chi^{4(1-\alpha)+i}}{4(1-\alpha)+i}.
\ee

Unlike the \citetalias{NFW97} profile, where the scale radius $\rs$ is the same as the radius at which the logarithmic density slope equals $-2$ (hereafter referred to as $r_2$), in a \citetalias{Dekel17} profile, the two radii are related by
$r_2 = [(2-\alpha)/1.5]^2 \rs$.
That is, the conventional concentration parameter, $c_2 = \rv / r_2$, is related to the \citetalias{Dekel17} concentration by
\be
\label{eq:concentration}
c_2 = \left(\frac{1.5}{2-\alpha}\right)^2 c.
\ee
The radius of peak circular velocity, $\rmax$, is related to $r_2$ by 
\be
\rmax = 2.25r_2 = (2-\alpha)^2 \rs.
\ee

The parameter $\alpha$ is the logarithmic density slope, $-\rmd\ln\rho/\rmd\ln r$, in the asymptotic limit $r\to0$, which may fall well outside the radial range of interest (for example between $0.01\rv$ and $\rv$). 
For the slope in the radial range of interest, the slope profile is given by
\be
s(r) = - \frac{\rmd\ln\rho}{\rmd\ln r} = \frac{\alpha+3.5\sqrt{x}}{1+\sqrt{x}}.
\ee
The slope at $0.01\rv$, widely used in the context of the cusp-core issue, is
\be \label{eq:s1}
s_{0.01} \equiv s(0.01\rv) = \frac{\alpha+0.35\sqrt{c}}{1+0.1\sqrt{c}}.
\ee
For $s_{0.01}$ values that are commonly seen in simulations and observations ($0-2$) and for a typical concentration (e.g., $c=10$), we have $\alpha\in(-1.11,1.53)$.
That is, $\alpha$ can be negative for realistic profiles (corresponding to a density that actually decreases towards the halo centre) and thus $s_{0.01}$ is a more physical quantity than $\alpha$ when it comes to comparing the cuspiness of density profiles. 

\subsubsection{\citet{MN75} disc profile}

A \citetalias{MN75} disc is specified by three parameters: the disc mass ($\Md$), a scale radius ($a$), and a scale height ($b$). 
The density and potential profiles are given by
\be
\label{eq:MNdensity}
\rho(R,z) = \frac{\Md b^2}{4\pi}\frac{aR^2+(a+3\zeta)(a+\zeta)^2}{\zeta^3[R^2+(a+\zeta)^2]^{5/2}}
\ee
and
\be
\label{eq:MNPotential}
\Phi(R,z) = -\frac{G\Md}{\sqrt{R^2 + (a+\zeta)^2}},
\ee
respectively, where $\zeta=\sqrt{z^2+b^2}$ and and $R$, $\phi$, and $z$ are the cylindrical coordinates.
For an axisymmetric disc whose distribution function only depends on $E$ and $L_z$, the radial and axial velocity dispersions are equal: $\sigma_R=\sigma_z\equiv \sigma$. 
Further assuming that the disc is an isotropic rotator, i.e., $\Vphibar^2/(\overline{\Vphi^2}-\sigma^2)=1$, we have $\sigma_\phi^2 = \overline{\Vphi^2} - \Vphibar^2 = \sigma^2$, and $\sigma^2$ is given by \citet{CP96} by
\be
\label{eq:MNVD}
\sigma^2(R,z) = \frac{G\Md^2 b^2}{8\pi\rho(R,z)}\frac{(a+\zeta)^2}{\zeta^2[R^2+(a+\zeta)^2]^3}.
\ee
The net rotation, $\Vphibar$, can therefore be expressed by
\be
\label{eq:MNrotation}
\Vphibar^2 = \Vc^2 + \frac{R}{\rho}\frac{\partial(\rho \sigma^2)}{\partial R} =\frac{G\Md^2 a b^2}{4\pi \rho}\frac{R^2}{\zeta^3[R^2+(a+\zeta)^2]^3},
\ee
where $\Vc^2(R,z) = R\partial \Phi/\partial R$ and $(R/\rho)\partial(\rho\sigma^2)/\partial R$  is the asymmetric-drift term.
\Eqs{MNVD} and (\ref{eq:MNrotation}) are useful for modeling dynamical friction (\se{orbit}).

\subsection{Initial conditions for satellite galaxies}
\label{sec:initial}

The initial conditions for a satellite galaxy include (1) the properties of the host system when the satellite enters the virial sphere, (2) the orbit of the incoming satellite, and (3) the DM, stellar, and gaseous properties of the incoming satellite. Here we describe them one by one.

\subsubsection{Initial host profile}
 
The host halo mass is known from the main branch (i.e., the branch that tracks the most massive progenitor) of the merger tree. To fully specify the host halo profile, we also need the structural parameter(s). The halo concentration can be obtained from an empirical relation calibrated via simulations \citep{Zhao09}, which relates the main branch merging history to the concentration parameter, $c_2$, by
\be
\label{eq:ConcentrationMAHrelation}
c_2(\Mv,z) = \left\{4^8 + \left[\frac{t(z)}{t_{0.04}(\Mv,z)}\right]^{8.4}\right\}^{1/8},
\ee
where $t(z)$ is the cosmic time at redshift $z$ and $t_{0.04}$ is the cosmic time when the host halo has assembled $4\%$ of its instantaneous mass, $\Mv(z)$, which we extract from the halo's merger tree as described in \se{tree}.
If the host system is only an \citetalias{NFW97} halo, then concentration and mass completely specifies it. 
For a more complicated setup, e.g., a \citetalias{Dekel17} halo with an embedded \citetalias{MN75} disc, one needs additional assumptions depending on the system of interest (see e.g., \se{application} for more details for MW/M31 analogues). 
The concentration $c$ and the slope $\alpha$ of a \citetalias{Dekel17} halo can be obtained from \eqs{concentration}, (\ref{eq:s1}), and (\ref{eq:ConcentrationMAHrelation}), with an assumption for $s_{0.01}$ that will be described in \se{HaloResponse}. 

\subsubsection{Initial orbit}

The initial orbit of a satellite can be specified by four pieces of information -- the location of virial-crossing, orientation of the orbital plane, orbital energy, and orbital circularity.
We assume that the infall locations are isotropically distributed on the virial sphere, and thus randomly draw an azimuthal angle ($\phi$) from $[0,2\pi]$ and a cosine polar angle ($\cos\theta$) from $[0,1]$. 
We parameterize the specific energy of an orbit, $E$, by a unitless parameter, $\xc=\rc(E)/\rv$, which is the radius of the circular orbit corresponding to the same orbital energy, $E$, in units of the virial radius of the host halo \citep[e.g.,][]{vdB17}. 
Orbital circularity, $\epsilon=j/\jc(E)$, is the ratio between the specific orbital angular momentum and that of a circular orbit of the same orbital energy.
We assume $\xc=1$, typical of cosmological orbits seen in simulations\footnote{To be more accurate, one can draw $\xc$ from orbital energy distributions extracted from simulations \citep[e.g.,][]{vdB17}, which show a median value around $\xc{\sim}1$. We opt to keep it simple and use $\xc=1$ in this work. After all, the correlation between initial orbital parameters and initial satellite properties is not clear yet. In an upcoming work (Green et al., in prep), we expand \SatGen to draw orbits according to a distribution extracted from cosmological simulations, following Li et al. (in prep). }
and draw $\epsilon$ from a distribution, $\rmd P/\rmd \epsilon = \pi\sin(\pi\epsilon)/2$, which approximates the $\epsilon$ distribution of infalling satellites measured in cosmological simulations \citep[e.g.,][]{Wetzel10, Jiang15, vdB17}. 

For orbit integration (\se{orbit}), we need to translate these orbital parameters ($\phi,\theta,\xc,\epsilon$) to the position vector, $\boldsymbol{r}$, and the velocity vector, $\boldsymbol{V}$.
Since \SatGen supports axisymmetric potentials, we work in the cylindrical coordinate system, i.e., $\boldsymbol{r}=(R,\phi,z)$ and $\boldsymbol{V}=(\VR,\Vphi,\Vz)$.
The initial speed at virial-crossing ($V$) is given by
\be
V = \sqrt{2[\Phi(\xc\rv) - \Phi(\rv)]+\Vc^2(\xc\rv)},
\ee
which is simply $\Vv$ for $\xc=1$.
Using the definition of $\epsilon$, we can derive the angle ($\tilde{\theta}$) between $\boldsymbol{V}$ and $\boldsymbol{r}$:
\be
\tilde{\theta} = \pi - \arcsin\left(\epsilon\xc \frac{\Vv}{V}\right).
\ee
In order to fully specify the orientation of the orbital plane, we need another angle for the velocity vector. 
We choose this angle to be the azimuthal angle ($\tilde{\phi}$) of $\boldsymbol{V}$ in the $\hat{\boldsymbol{\theta}}$-$\hat{\boldsymbol{\phi}}$-$\hat{\boldsymbol{r}}$ frame, and draw $\tilde{\phi}$ randomly from $[0,2\pi]$.
Finally, we can specify all the phase-space coordinates of the infalling satellite:
\bad
R & = \rv\sin\theta,\\
\phi & = \phi, \\
z & = \rv \cos\theta,\\
\VR & = V (\sin\tilde{\theta}\cos\tilde{\phi}\cos\theta + \
            \cos\tilde{\theta}\sin\theta) ,\\
\Vphi & = V \sin\tilde{\theta}\sin\tilde{\phi},\\
\Vz & = V (\cos\tilde{\theta}\cos\theta -\sin\tilde{\theta}\cos\tilde{\phi}\sin\theta).
\ead

\subsubsection{Initial subhalo density profiles}
\label{sec:HaloResponse}

In cosmological $N$-body simulations, halo density profiles are well-approximated by \citetalias{NFW97} profiles. Therefore, if \SatGen is used to emulate an $N$-body simulation, in order to initialize a subhalo profile we only need to compute the concentration parameter $c_2$ using \eq{ConcentrationMAHrelation}.

To emulate hydro-simulations, we need to account for the fact that haloes react to baryonic processes that cause their profiles to deviate from \citetalias{NFW97}. 
The {\it halo response} to baryonic processes is mass-dependent \citep[e.g.,][]{DiCintio14a, Dutton16, Tollet16, Freundlich20}: qualitatively, low-mass haloes ($\la10^{11}\Msun$) are susceptible to supernovae-driven gas outflows, becoming less concentrated and developing a flatter core; in contrast, massive haloes ($>10^{12}\Msun$) tend to contract as cold gas condenses in the centre, becoming cuspier. 
The halo response strength depends on the sub-grid physics adopted in the simulations.
This is especially relevant for massive dwarf galaxies ($\Mv\sim10^{10.5}\Msun$). 
Notably, simulations featuring bursty star formation, and thus strong episodic supernovae outflows, yield a strong halo response, whereas simulations with smooth, continuous star formation exhibit a negligible halo response in the dwarf regime \citep{Bose19, Dutton19}. 
The nature of the star formation burstiness, and thus the strength of the halo response, is closely related to the sub-grid recipe for star formation and is still highly uncertain and under debate.

Following \citet{DiCintio14a, DiCintio14b} and \citet{Tollet16}, we parameterize the halo response with two relations: (1) the ratio of the hydro-simulation concentration and the corresponding DM-only concentration, $c_2/c_{\rm 2,DMO}$ as a function of the stellar-to-halo-mass ratio (SHMR), $X = \Ms/\Mv$, and (2) the logarithmic DM density slope measured at ${\sim}1\%$ of the virial radius, $s_{0.01}$, as a function of the SHMR. 
Specifically, the concentration ratio can be expressed by 
\be
\label{eq:ConcentrationSHMRrelation}
\frac{c_2}{c_{\rm 2,DMO}} =  a_0 + a_1 X^{b_1} - a_2 X^{b_2},
\ee
where the constants $a_i$ and $b_i$ are simulation-specific and are chosen according to the simulation that one wishes \SatGen to emulate. 
For example, we find that $(a_0,a_1,a_2)=(1.14, 186, 1)$ and $(b_1,b_2)=(1.37,0.142)$ describe the halo response of the NIHAO \citep{Wang15} simulations accurately \citep{Freundlich20}. 
For these parameters, $c_2 / c_{\rm 2,DMO}$ approaches unity at $\Ms/\Mv<10^{-4}$, where star formation is weak and feedback effects are minimal (typical of low-mass haloes), is less than unity ($\sim0.7$) at $\Ms/\Mv\sim10^{-2.5}$ (typical of massive dwarf galaxies where feedback effects are maximal), and becomes $>1$ at $\Ms/\Mv>10^{-2}$ (where adiabatic contraction dominates).
Similarly, the inner density slope $s_{0.01}$ can be expressed as
\be
\label{eq:SlopeSHMRrelation}
s_{0.01} \equiv -\frac{\rmd\ln\rho}{\rmd\ln r}|_{0.01\rv} =  \log\left[n_1 \left(1+\frac{X}{X_1}\right)^{-\xi_1} + \left(\frac{X}{X_0}\right)^{\xi_0}\right] + n_0,
\ee
where the constants $X_i$, $n_i$, and $\xi_i$ are, again, chosen to reflect the simulation sub-grid physics of interest \citep{Tollet16}.
For the NIHAO simulations, \citet{Freundlich20} find that $(n_0,n_1) =(1.45,1)$, $(\xi_1,\xi_0) =(2.14,0.21)$, and $(X_0,X_1)=(2.54\times10^{-3},9.87\times10^{-4})$.
This describes the phenomenon that DM cores form if $X\sim10^{-3}$-$10^{-2}$, cusps remain present for smaller $X$, and baryons deepen the gravitational potential at larger $X$. 
We add random Gaussian noise with $\sigma=0.1$ and 0.18 to the $c_2/c_{\rm 2,DMO}$ and $s_{0.01}$ values, respectively, based on \citet{Freundlich20} and \citet{Tollet16}. 
We note that the aforementioned halo response is likely quite generic for simulations featuring bursty star formation and episodic strong feedback, such as the FIRE simulations \citep{Hopkins14, Hopkins18}. 

We use the \citetalias{Dekel17} profile to describe subhaloes affected by feedback. 
From \eq{DekelDensity}, we can show that the slope at $r\to 0$ ($\alpha$) and the slope at $r=0.01\rv$ ($s_{0.01}$) are related by
\be
\label{eq:slope}
\alpha = s_{0.01} (1+0.1\sqrt{c}) - 0.35\sqrt{c}. 
\ee
Using \eqs{concentration}, (\ref{eq:ConcentrationMAHrelation}), (\ref{eq:SlopeSHMRrelation}), (\ref{eq:slope}), and a SHMR, we can completely specify a \citetalias{Dekel17} subhalo at infall.
\footnote{For \citetalias{Einasto65} profiles, an expression analogous to \eq{slope} between the \citetalias{Einasto65} shape index and $s_{0.01}$ can be derived. See \app{profiles} for details.}

We emphasize that one of the goals of \SatGen is to quantify the influence of different halo response models on satellite statistics, and thus to distinguish the underlying sub-grid recipes adopted in simulations using observed satellite statistics. 
More specifically, the logic is the following. 
On the theory side, while it is computationally expensive to run simulations with adequate resolution for studying satellite galaxies, it is relatively cheap to simulate a suite of field galaxies that cover a wide range in mass and SHMR. 
These types of simulation suites, e.g., FIRE/FIRE-II \citep{Hopkins14, Hopkins18}, NIHAO \citep{Wang15}, APOSTLE \citep{Sawala15}, and Auriga \citep{Grand17}, provide us with halo response templates, $(c_2/c_{\rm 2, DMO})(X)$ and $s_{0.01}(X)$ \citep[e.g.,][]{Tollet16, Bose19}, which are used as inputs for the \SatGen model.
\SatGen then propagates the difference in halo response to satellite structures because, as will be detailed in \se{tracks}, satellites of different initial structures evolve differently in response to tidal effects. 
In this way, \SatGen produces satellites as would be produced by high-resolution simulations using the corresponding sub-grid recipe. 
On the observational side, galaxy structure and halo structure measurements are usually performed on galaxies of known distances, which are typically satellites. 
By propagating the baryonic effects obtained from zoom-in simulations of centrals onto satellite populations, \SatGen facilitates the comparison between theory and observation. 

\subsubsection{Initial baryonic properties}
\label{sec:InitialSize}

Apart from subhalo properties, we also model the stellar mass, stellar size, and gas distribution.  
We assign a stellar mass to an infalling satellite using the SHMR from halo abundance matching. 
In particular, we use the expression of stellar mass ($\Ms$) as a function of halo mass ($\Mv$) and redshift $z$ by \citet{RP17}, assuming a scatter of 0.15 dex in $\Ms$ at a given $\Mv$.
Abundance matching also provides insight on how the galaxy size is related to the host halo structure -- \citet{Kravtsov13} and \citet{Somerville18} found that galaxy size scales linearly with host halo virial radius, $\reff \sim 0.02\rv$, insensitive to morphology.
\citet{Jiang19a} verified this relation in two different suites of cosmological hydro-simulations, finding that the proportionality constant does not reflect halo spin but strongly correlates with halo concentration, $c_2$. In particular, 
\be
\label{eq:size}
\reff = 0.02 (c_2/10)^{-0.07} \rv.
\ee
The dependence on halo concentration introduces a redshift and assembly history dependence into the galaxy size.
We adopt this relation in order to initialize the satellite's stellar size, assuming a log-normal scatter with $\sigma=0.15$ dex in $\reff$ at fixed $\rv$, as found by \citet{Jiang19a}. 
Note that we track the evolution in the satellite's stellar half-mass radius without making any specific assumptions about the underlying density profile of the stars.

Following \citet{Zinger18}, we assume that the circumgalactic medium (CGM) of a galaxy is in hydrostatic equilibrium with the host halo and, to a good approximation, follows the halo profile according to
\be
\label{eq:GasProfile}
\rhog(r) = \fg \rho(r), 
\ee
where $\fg$ is the ratio of the total CGM gas mass to virial mass. 
For incoming satellites, we can write
\be
\fg = \frac{\fb}{1-\fb} - \frac{\Ms}{\Mv},
\ee
where the baryonic fraction, $\fb$, is given by \citet{Okamoto08} as
\be
\label{eq:BaryonicFraction}
\fb(\Mv,z) = \frac{\Omega_{\rm b}}{\Omega_{\rm m}}\left\{ 1 + 0.587\left[\frac{\Mv}{\Mc(z)}\right]^{-2}\right\}^{-3/2},
\ee
where $\Mc(z)$ is the mass below which galaxies are strongly affected by photoionization. 
We adopt $\Mc(z)$ from the numerical values given by \citet{Okamoto08}.
This recipe implicitly assumes that supernovae feedback does not remove hot gas from the halo.

The prescriptions in \S\ref{sec:HaloResponse} and \S\ref{sec:InitialSize} apply both to the central host and to the satellites at the moment of infall.

\subsection{Orbit integration and dynamical friction}
\label{sec:orbit}

We follow the orbits by treating satellites as point masses. 
At each timestep, \SatGen solves the equations of motion in the cylindrical frame using an order 4(5) Runge-Kutta method.\footnote{We use the `dopri5' integrator as implemented in {\tt scipy.integrate.ode}.}
We solve
\be
\label{eq:EOM}
\ddot{\boldsymbol{r}} = -\nabla \Phi\ + \aDF,
\ee
where $\boldsymbol{r}=(R,\phi,z)$ is the position vector, $\Phi$ is the gravitational potential, and $\aDF$ is the acceleration due to dynamical friction (DF), which is modeled using the \citet{Chandrasekhar43} formula,
\be
\label{eq:DF}
\aDF = -4\pi G^2 m \sum_{i} \lnL_i\, \rho_i(\boldsymbol{r}) F(<V_{{\rm rel},i})\frac{\boldsymbol{V}_{{\rm rel},i}}{V_{{\rm rel},i}^3}\,.
\ee 
Here the summation is over all of the components of the host system (e.g., $i=$halo, disc, and bulge, following \citealt{TB01} and \citealt{Penarrubia10}), $m$ is the instantaneous satellite mass, $\lnL_i$ is the Coulomb logarithm, $\boldsymbol{V}_{{\rm rel},i}$ is the relative velocity of the satellite with respect to the streaming motion of the particles of component $i$, and $F(<V_{{\rm rel},i})$ is the fraction of local host particles contributing to DF.
For simplicity, we assume that the velocity distributions of all of the host components are Maxwellian and isotropic such that 
\be
\label{eq:VelocityDistribution} 
F(<V_{{\rm rel},i})= \erf(X_i) - \frac{ 2 X_i }{\sqrt{\pi}}e^{-X_i^2},
\ee
where $X_i\equiv V_{{\rm rel},i} / (\sqrt{2}\sigma_i)$, with $\sigma_i(\boldsymbol{r})$ the one-dimensional velocity dispersion of component $i$. 
\footnote{In principle, for a composite potential in Jeans equilibrium and with isotropic velocity distribution, the ``one-dimensional velocity dispersion of component $i$'' ($\sigma_i$) is not well-defined, because the velocity dispersion should be calculated as a quantity for the whole system using the Jeans equation, which gives (e.g., for spherical systems): $\sigma^2(r) = G/[\sum_i\rho_i(r)]\int_\infty^r \sum_i\rho_i(r^\prime)[\sum_i M_i(r^\prime)/r^{\prime 2}] \rmd r^\prime \ga \sigma_i^2(r)$. However, in practice, we find that using the $\sigma_i$ of each component as if they were in equilibrium separately in isolation yields little difference in terms of the rate of orbital decay compared to using the overall $\sigma(r)$. This is mainly because $V_{{\rm rel},i}$ is usually larger than $\sigma(r)$, so $F(<V_{{\rm rel},i})$ is often not far from its maximum value of unity. Additionally, satellite mass loss and the choice of $\lnL$ both have larger impacts on DF than the detailed choice of $\sigma$. Therefore, we opt to use the $\sigma_i$ of individual components, following \citet{TB01}.}

The Coulomb logarithm and the relative velocity depend on the host component of interest. 
For spherical components such as the halo or bulge, we adopt 
$\lnL_i = \xi \ln(M_i/m)$,
where the factor $\ln(M_i/m)$ is a widely used form for the Coulomb logarithm \citep[e.g.,][]{Gan10}, with $M_i$ and $m$ the host mass and satellite mass, respectively, and $\xi$ a fudge factor that accounts for the weakening of orbital decay when the density profile is cored \citep[e.g.,][]{Read06b}.
Orbital decay becomes completely stalled where the host density profile is flat, i.e., if $s=-\rmd \ln\rho/\rmd\ln r=0$, whereas orbital decay continues where the profile is cuspy, i.e., if $s\ga1$. 
For simplicity, we assume $\xi =\min(s,1)$.
For discs, we use $\lnL=0.5$, following \citet{Penarrubia10}.

For spherical components, we use the orbital velocity $\boldsymbol{V}$ for $\boldsymbol{V}_{{\rm rel},i}$; i.e., we ignore the net spin of a halo or a bulge.  
Discs, however, have net rotation, so we use $\boldsymbol{V}_{\rm rel,d} = \boldsymbol{V} - \Vphibar \hat{\boldsymbol{\phi}}$, where the mean rotation $\Vphibar$ is given by \eq{MNrotation}.

We caution that our DF treatment is only approximate, and, as with any other attempt of modeling subhalo orbit with the \citet{Chandrasekhar43} formula, it carries a few conceptual inaccuracies. 
For instance, the \citet{Chandrasekhar43} formula assumed point masses moving in medium of uniform density, whereas a subhalo has an extended mass distribution and the host density along its orbit is not constant.
The aforementioned choices of the Coulomb logarithm are therefore empirical corrections when extending the formula to applications beyond its assumptions.
More fundamentally, \citet{Chandrasekhar43} considers DF to be a local effect due to the trailing gravitational wake, while DF is actually a global effect due to a response density that can operate at long distances \citep[e.g.,][]{Weinberg89}.
However, we have verified that the impact on satellite statistics due to this approximation is rather limited.
Notably, for the experiments in \se{application}, we found that setting the disc DF term to zero only yields a $\sim1\%$ increase in the number of surviving satellites, and changing the whole $\aDF$ by a factor of two results in only a $\sim$10\% change in the abundance of satellites.

\subsection{Tidal stripping and ram pressure stripping}
\label{sec:stripping}

Satellites lose DM mass and stellar mass to tides, and they lose gaseous mass to ram pressure when their orbits bring them close enough to the centre of the host system. 

We estimate the instantaneous tidal radius of the satellite, $\lt$, at each point along its orbit by solving
\be
\label{eq:TidalRadius}
\lt = r \left[ \frac{m(\lt) / M(r)}{2-\frac{\rmd\ln M(r)}{\rmd\ln r}+\frac{\Vt^2(\boldsymbol{r})}{\Vc^2(r)} } \right]^{1/3}
\ee
\citep[e.g.,][]{King62, TB01, ZB03}, where $m(l)$ and $M(r)$ are the enclosed mass profiles of the satellite and host, respectively, and $\Vt(\boldsymbol{r})= |\hat{\boldsymbol{r}} \times \boldsymbol{V}|$ is the instantaneous tangential speed.
The first two terms in the denominator represent the gravitational
tidal force -- obviously, tidal stripping depends on the local mass profile of the host (see \citealt{Dekel03} for a thorough discussion).  
The third term represents the differential centrifugal force across the satellite due to its orbital motion about the halo centre.

Although the tidal radius is widely used to model tidal stripping, it is an ill-defined concept for several reasons \citep[e.g.,][]{vdB18}.
For example, the Hill surface is not spherical or infinitesimally thin \citep{Read06a,Tollet17}. 
Because of this, we express the instantaneous mass loss rate as
\be
\label{eq:MassLossRate}
\dot{m} = -\mathcal{A} \frac{m(>\lt)}{\tdyn(r)},
\ee   
where we have introduced a fudge parameter $\mathcal{A}$ as the stripping efficiency to incapsulate uncertainties in the definition of the tidal radius.
As such, the timescale on which stripping occurs is the local dynamical time $\tdyn(r) = \sqrt{3\pi/16G\rhobar(r)}$ divided by $\mathcal{A}$ (with $\rhobar(r)$ the average density of the host system within radius $r$, including the baryonic components).
We calibrate the mass loss rate model using high-resolution idealized simulations and find $\mathcal{A}\approx0.55$ (Green et al. in prep).
\footnote{
In several previous studies \citep[e.g.,][]{ZB03, Zentner05, Pullen14, vdB18}, the stripping time is assumed to be the instantaneous orbital time divided by a fudge factor, i.e., $(2\pi r/\Vt)/A$, with $A=1-6$ across the studies. Our choice of $\mathcal{A}=0.55$ corresponds roughly to $A{\sim}1.65$ for a typical cosmological orbit, bracketed by literature values but on the inefficient-stripping end. The stripping efficiency parameter may weakly depend on the density profile used for describing the evolved subhaloes.}
The mass evolution over a timestep $\Delta t$ is then given by
\be
\label{eq:MassLoss}
m(t+\Delta t) = m(t) + \dot{m} \Delta t .
\ee

Similarly, if a higher-order satellite (see \fig{cartoon} for definition) stays outside the tidal radius of the hosting satellite for more than a time of $\tdyn(l)/\mathcal{A}$, where $\tdyn(l)$ is the local dynamical time of the hosting satellite, it is released to the lower-order host, picking up a new orbital velocity that is the superposition of its velocity with respect to the previous hosting satellite and the velocity of the hosting satellite with respect to the lower-order host. 

Analogous to how the tidal radius is defined, a ram pressure radius ($\lRP$) can be defined as the satellite-centric distance where the self-gravitational restoring force per unit area balances the ram pressure exerted by the gaseous host halo. 
We compute $\lRP$ at each point along the orbit by solving
\be
\label{eq:RamPressure}
\kappa \frac{Gm(\lRP)\rhog(\lRP)}{\lRP} = \rhog(\boldsymbol{r})V(\boldsymbol{r})^2,
\ee
where $\kappa$ is a factor of order unity \citep[][$\kappa=0.5-2$, depending on assumptions made in calculating the gravitational restoring force]{Zinger18}, and we take for simplicity $\kappa=1$.
The mass loss rate of the gaseous halo is given by
 \be
\label{eq:GasLossRate}
\dot{m}_{\rm gas}= - \frac{\mg[>\min(\lt,\lRP)]}{2\tdyn(r)}.
\ee  
In practice, $\min(\lt,\lRP)=\lRP$ in most cases, i.e., ram pressure stripping is usually more efficient than tidal stripping for gas.

\subsection{Evolution of satellite structure}
\label{sec:tracks}

Satellites react to two competing tidal effects: tidal stripping, which takes mass away and makes satellite smaller, and tidal heating, which injects orbital kinetic energy into the satellite, causing it to expand.
While tidal stripping can be analytically estimated (\se{stripping}), the effect of heating, or the net structural response to both tidal effects, is not easily captured by analytical arguments. 
Several studies have resorted to using idealized simulations to tabulate satellite structural evolution due to the tidal field as a function of the mass that has been lost \citep{Hayashi03, Penarrubia08, Penarrubia10, Errani15, Errani18, GB19}.\footnote{But see also Du et al. (in prep), which studies the tidal heating of subhaloes using idealized $N$-body simulations and derives analytical formulae that accurately approximate the effects of tidal heating on subhalo density profiles.}
Notably, \citet{Hayashi03} and \citet{Penarrubia08, Penarrubia10} found that subhalo density profiles depend solely on the density profile at infall and the total amount of mass lost thereafter.
In particular, they describe the evolution of the maximum circular velocity ($\vmax$) and the radius at which the circular velocity reaches the maximum ($\lmax$) using a generic function,
\be
\label{eq:DMTracks}
g(x) = \left(\frac{2}{1+x}\right)^{\mu} x^{\eta},
\ee 
where $g(x)=\vmax(t)/\vmax(0)$ or $\lmax(t)/\lmax(0)$, $x$ is the bound mass fraction ($m(t)/m(0)$), and $\mu$ and $\eta$ are the best-fit parameters calibrated against idealized simulations.
\citet{Penarrubia10} found that $\mu$ and $\eta$ depend on the initial inner logarithmic density slope of the satellite, $s_{0.01}$ (see \app{tracks} for their values). 
These relations, also known as {\it tidal-evolution tracks}, are scale-free, independent of the orbital parameters, and only marginally sensitive to the initial concentration of the subhaloes \citep{GB19}, which we ignore here.

\citet{Errani18} extended tidal tracks to describe the evolution of the stellar mass ($\ms$) and half-stellar-mass radius ($\leff$). 
In particular, they found that
\be
\label{eq:StellarTracks}
\tilde{g}(x) = \left(\frac{1+\xs}{x+\xs}\right)^\mu x^\eta,
\ee 
where $\tilde{g}(x)=\ms(t)/\ms(0)$ or $\leff(t)/\leff(0)$ and $x=\mmax(t)/\mmax(0)$, with $\mmax$ the subhalo mass within the maximum-circular-velocity radius, $m(\lmax)$. 
Here, the parameters, $\mu$, $\eta$, and $\xs$, depend not only on the initial density slope, $s_{0.01}(0)$, but also on how compact the stellar component initially is with respect to the hosting subhalo, measured by $\leff(0)/\lmax(0)$.
Note that by using these tidal tracks, we do not assume density profiles for stellar mass or explicitly model tidal stripping of stars; instead, we updated the evolved stellar mass and half-mass radius assuming that they are coupled to the evolution of the subhaloes through $m(\lmax)$.  
We list the parameter values in \app{tracks}, but summarize the tidal tracks qualitatively here as follows: satellite size generally increases with subhalo mass loss, which manifests due to tidal heating and the re-virialization response to tidal stripping and heating; only cuspy satellites ($\alpha \ga 1$) can become more compact, and the size decrease occurs only after significant subhalo mass loss.

With the tidal tracks described by \eqs{DMTracks}-(\ref{eq:StellarTracks}), the formula for tidal stripping, \eqs{TidalRadius}-(\ref{eq:MassLoss}), and the initial profile as set up in \se{initial}, we can completely specify the {\it evolved} subhalo profile, the stellar mass, and the stellar size at each timestep along the orbit. 
For this, a conversion between $\vmax$ and $\lmax$ and the parameters that are directly used to define a subhalo density profile, e.g., the concentration $c$ and overdensity $\Delta$, is needed.
We provide details on such a conversion in \app{tracks}.
For the gas distribution, we assume that the remaining gas follows the evolved subhalo profile as in \eq{GasProfile}, with $\fg = \mg(t) / m(t)$.

\subsection{Improvements compared to previous models}
\label{sec:improvement}

\SatGen combines the wisdom of earlier models and improves in important ways.
Most previous models have focused on DM subhaloes \citep{TB01, ZB03, Zentner05, Gan10, Penarrubia10, JB16}, whereas \SatGen takes baryonic properties into consideration. 
A couple of models have included certain details of baryonic processes \citep{Carleton19, Nadler19}, but \SatGen is more thorough. 

For example, the model by \citet{Nadler19} considers the stellar component.
It initializes the satellite stellar size in the same way as \SatGen, but for the size evolution it only considers size decrease due to tidal stripping and neglects expansion due to tidal heating, which is a process that is essential for producing UDGs in dense environments \citep{Carleton19, Jiang19b}. 
Also, tidal stripping in this model is treated in an orbit-averaged sense, as in \citet{vdB05} and \citet{JB16}.
This treatment washes out detailed mass and structural evolution along the orbits. 

The model by \citet{Carleton19} uses the same tidal tracks as used in \SatGen; however, it applies abrupt tidal truncation to satellites at orbital pericentres such that pericentres are the only locations where the satellites lose mass.
This is not accurate for circular orbits or any orbits with $\epsilon \ga 0.5$.
In addition, the \citet{Carleton19} model relies on cosmological $N$-body simulations for merger trees, orbits, and initial conditions.
In contrast, \SatGen can generate larger samples using the EPS formalism, which is useful for studying the halo-to-halo variance of satellite properties, and can follow the orbits self-consistently.

\subsection{Illustration and workflow}
\label{sec:workflow}

We present an idealized example of a massive satellite orbiting a MW-sized halo in Appendix \ref{app:illustration} in order to provide an intuitive illustration (\fig{test_evolve}) of the orbit integration and satellite evolution prescriptions described in \se{profiles}-\se{tracks}.

When using \SatGen for a cosmological setup, we summarize the workflow as follows: 
\begin{itemize}[leftmargin=*]
\item[1.] Starting with a target halo of a given mass and redshift, draw halo merger trees according to \se{tree}.
\item[2.] Initialize host and satellite properties according to \se{initial}, using density profiles introduced in \se{profiles} and Appendix \ref{app:profiles}, and considering halo response models that are characteristic of certain cosmological hydro-simulations.
\item[3.] Evolve the satellites: integrate the orbit according to \se{orbit} and update the masses and profiles of the satellites and the host for every timestep of $\Delta t = 0.1\tdyn(z)$, according to \se{tracks}. 
\end{itemize}
This procedure is somewhat similar to that of zoom-in simulations, in the sense that both \SatGen and zoom-in simulations start with a target halo and then trace the progenitors back in time, finally evolving forward in time to refine the small-scale structures. 

\section{Satellites of MW/M31 sized host haloes}
\label{sec:application}

\begin{figure*}
	\includegraphics[width=\textwidth]{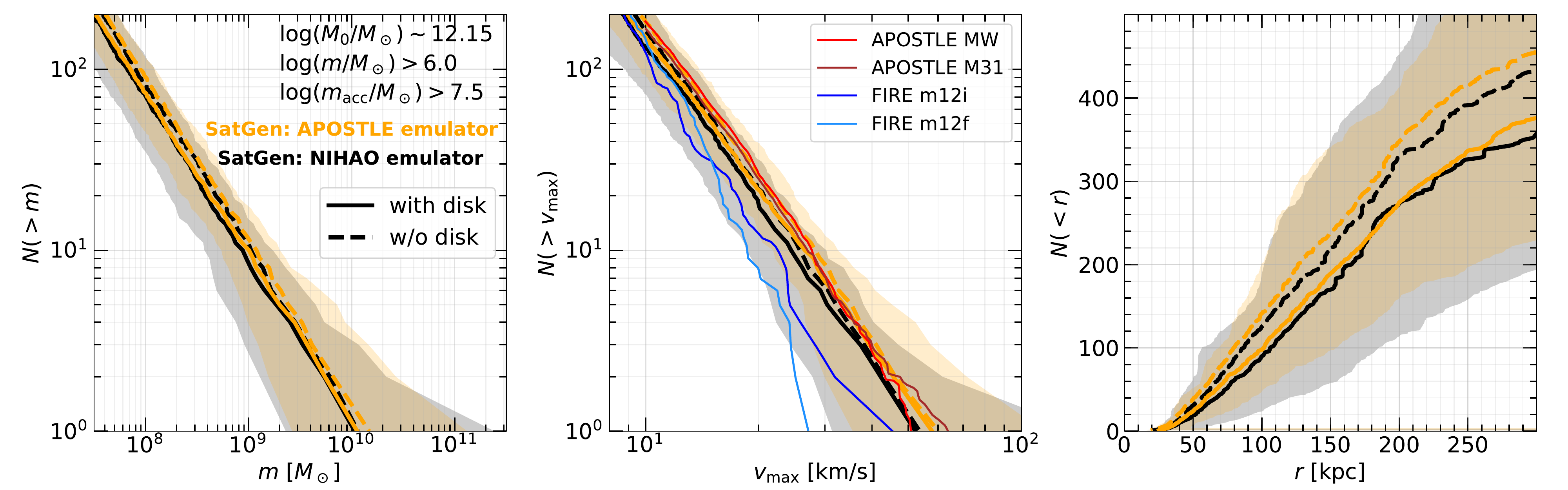}
    \caption{Satellite statistics predicted by \SatGen in NIHAO- and APOSTLE-emulating modes -- the cumulative subhalo mass function $N(>m)$ ({\it left}), subhalo $\vmax$ function $N(>\vmax)$ ({\it middle}), and radial distribution $N(<r)$ ({\it right}) of all of the surviving satellites in MW/M31-sized hosts (where ``surviving'' means $m>10^6\Msun$ at $z=0$ and ``MW/M31-sized'' means that the present-day host halo mass is in the range $M_0=10^{12-12.3}\Msun$; see \se{SHMF} for details). 
    Thick lines represent the median model predictions, with solid and dashed lines differentiating the cases with and without a disc potential. 
    The colors differentiate results from the NIHAO emulator (black) and the APOSTLE emulator.  
    Shaded bands indicate halo-to-halo variance (3-97 percentiles). 
    The thin lines in the middle panel are APOSTLE and FIRE simulation results for the $\vmax$ function \citep{Sawala15, GK17}. 
    Halo response differences result in a relatively minor effect: the NIHAO-like feedback yields $\sim5\%$ fewer satellites than the APOSTLE-like model. 
    A baryonic disc reduces the abundance of surviving satellites within 300 (100) kpc by $\sim20\%$ (30\%).
    Both baryonic effects are weak compared to the halo-to-halo variance. 
    }
    \label{fig:statistics_all}
\end{figure*}

For a proof-of-concept application, we use \SatGen to generate satellite galaxies for MW/M31-sized host systems, studying baryonic effects on satellite statistics including subhalo abundance, spatial distribution, and internal structures. 
In particular, we highlight the impact of two separate baryonic effects. The first is the impact that (supernova) feedback can have on the central density profile of the (sub)haloes hosting satellites. We refer to this as the {\it internal} effect due to baryons. The second is the impact that the baryonic disc of the host system has on the orbital and tidal evolution of satellites. In what follows we refer to these as the {\it internal} and {\it external} baryonic effects, respectively.

\subsection{Model setup and satellite statistics}
\label{sec:SHMF}

\begin{table}
\caption{Halo response relations adopted by the two simulation emulators considered in \se{application}.}
 \begin{threeparttable}
 \centering
\begin{tabular}{lll} 
\hline
\hline
  & NIHAO emulator \tnote{a} & APOSTLE emulator \\
\hline
\eq{ConcentrationSHMRrelation} & {\it for concentration} &\\
$a_0$   &  1.14     &  1  \\
$a_1$   &   186   & 186     \\
$a_2$   &   1   & 0     \\
$b_1$   &   1.37   &  --    \\
$b_2$   &   0.142   &  --    \\
\hline
\eq{SlopeSHMRrelation} & {\it for inner density slope} &\\
$n_0$   &  1.45 &  1.45  \\
$n_1$   &   1  & 1     \\
$X_0$   &   $2.54\times10^{-3}$   & $2.54\times10^{-3}$     \\
$X_1$   &   $9.87\times10^{-4}$  &  --    \\
$\xi_0$  &    0.21   &  0.21    \\
$\xi_1$  &    2.14   & 0    \\
\hline
\end{tabular}
\begin{tablenotes}
\item[a]  \citet{Freundlich20}.
\end{tablenotes}
\end{threeparttable}
\label{tab:HaloResponse}
\end{table}

We consider two different halo response models, which are representative of simulations of bursty star formation and strong supernovae feedback, such as NIHAO \citep{Wang15} and FIRE \citep{Hopkins14, Hopkins18}, and of simulations of non-bursty star formation and weaker feedback, such as APOSTLE \citep{Sawala15} and Auriga \citep{Grand17}.
We denote these two models as the NIHAO emulator and APOSTLE emulator, respectively, and tabulate the parameters of their halo response curves, as in \eqs{ConcentrationSHMRrelation}-(\ref{eq:SlopeSHMRrelation}), in Table \ref{tab:HaloResponse}. 

For each emulator, we randomly generate 100 merger trees for MW- and M31-sized haloes ($\Mv=10^{12-12.3}\Msun$ at $z=0$), recording progenitor haloes down to $10^{7.5}\Msun$ up to $z=20$.
We initialize the satellites and hosts as described in \se{initial} -- at this stage, the halo response relations are taken into account.\footnote{For this proof-of-concept study, we opt to only follow the DM and stellar components, ignoring the gaseous components.}
We then evolve the satellites, considering two cases.
In one case, the host potential is just a DM halo following the \citetalias{Dekel17} profile, as determined by the merger tree and the initialization procedure. 
In the other case, the host potential consists of both the DM halo and a galactic disc.
The disc mass is set to be $0.1$ times the instantaneous halo mass, i.e., $\Md(z)=0.1\Mv(z)$.
The disc follows a \citetalias{MN75} profile with $b/a=1/25$.
The disc size, $a$, is determined using the half-mass radius, $\reff$, as given by \eq{size}, and the relation between the \citetalias{MN75} $a$ and $\reff$, as given by \eq{MNsize}. 
Our discs are similar to those of \citet{Penarrubia10} in terms of mass and axis ratio. 
While approximately mimicking the cold discs of the MW or M31, these parameters are chosen mainly for illustration purposes and are not intended to reproduce the actual discs in the MW or M31 in any detail.  In fact, they are on the massive side of the observationally-inferred values \citep[e.g.,][]{Sofue13}.

In total we have four suites of simulations for a total of 400 MW/M31 sized haloes -- we have two suites for each simulation emulator and, for each emulator, we consider the case with and without the embedded galactic disc.
The merger trees and initial satellite structures of the with-disk and no-disk models are identical.
This enables us to quantify the disc effect.

\fig{statistics_all} presents the cumulative subhalo mass functions, $N(>m)$, subhalo $\vmax$ functions, $N(>\vmax)$, and satellite galactocentric-distance distributions, $N(<r)$, for all of the surviving satellites in the four suites at $z=0$. 
Here, we define ``surviving'' as having subhalo mass larger than $10^6\Msun$ and have verified that our results are not sensitive to this arbitrary mass threshold. 
Lines represent the median mass, $\vmax$, or distance at fixed number $N$, and the shaded bands indicate the 3-97 percentiles, reflecting the halo-to-halo variance due to random assembly histories.  
We overplot the $\vmax$ functions from the FIRE and APOSTLE simulations, finding that the \SatGen predictions are in reasonable agreement with the simulation results. 
We emphasize that this agreement is achieved without tuning any of the model parameters. 
We think that given the differences among the simulations, the halo-to-halo scatter, and the concern on the reliability of the simulation results due to numerical disruption \citep{vdB18}, there is no need to fine-tune the model to match the simulations in detail. 

The census of bright satellites ($\ms>10^5\Msun$) of MW and M31 is relatively complete \citep[e.g.,][]{Tollerud08}, so we use them as our observational benchmarks. 
\fig{statistics_massive} presents the \SatGen $\vmax$ functions and radial distributions for the massive surviving satellites with $\ms>10^5\Msun$ at $z=0$, and compares them with those of the \citet{McConnachie12} observational sample of MW/M31 satellites.  
We find that the model predictions agree well with those of the actual MW/M31 satellites.
Notably, the median radial distribution from the NIHAO emulator agrees with the MW and M31 observations at percent-level out to $\sim150$ kpc from the galactic centre, and even the observational results at the outskirts are well within the halo-to-halo variance of the model predictions.

\subsection{Effects of different baryonic physics}
\label{sec:DifferentHaloResponse}

\begin{figure*}
	\includegraphics[width=0.77\textwidth]{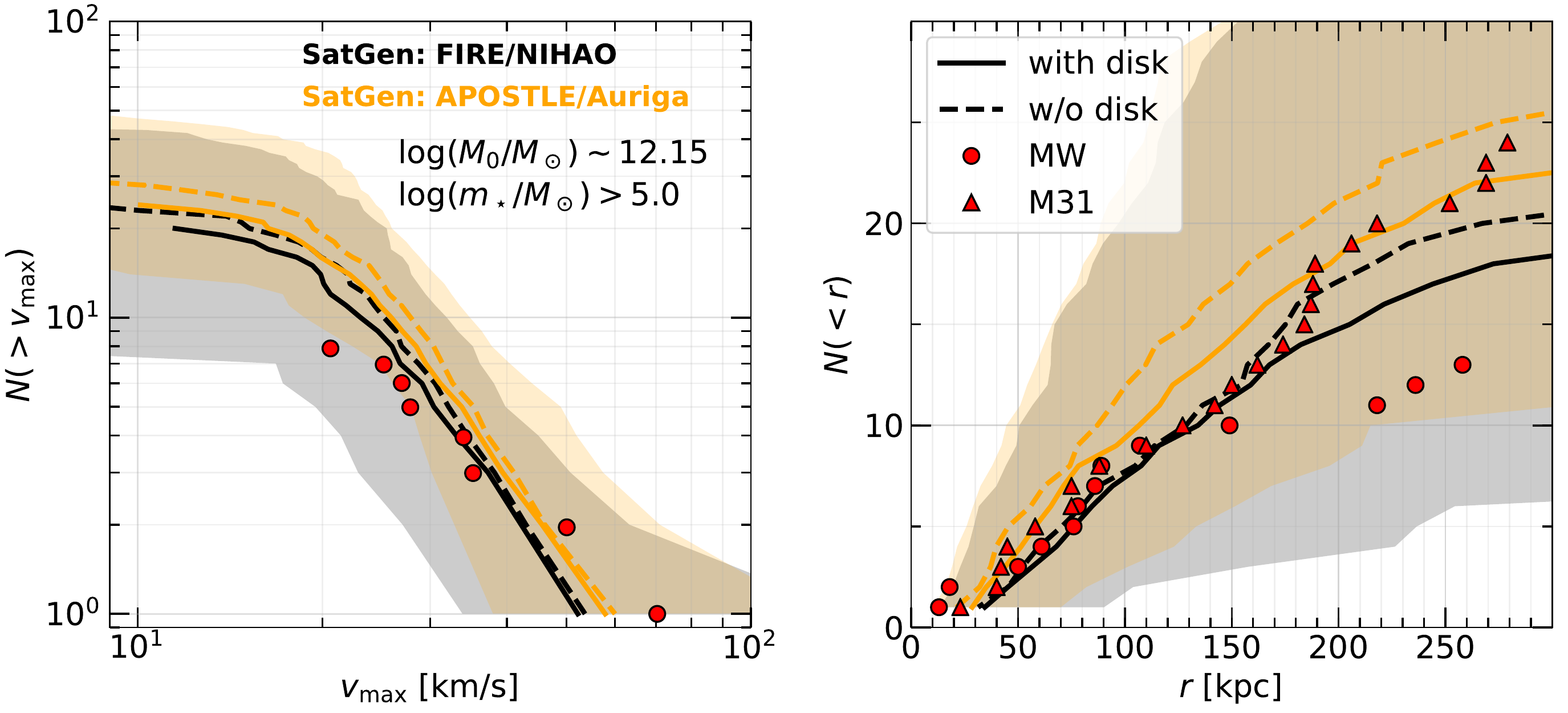}
    \caption{Subhalo $\vmax$ functions and radial distributions of massive satellites ($\ms>10^5\Msun$), comparing model predictions for MW/M31-sized host haloes (lines) and observations of the actual MW/M31 (symbols). 
    The shaded areas indicate halo-to-halo variance (3-97 percentiles, for the models with discs). 
    The flattening of the $\vmax$ function at the low-mass compared to the middle panel of \fig{statistics_all} is simply due to the stellar mass cut.
    The APOSTLE-like feedback on average yields ${\sim}25\%$ more massive satellites than the NIHAO-like feedback, illustrating that cuspier and denser satellites are more resistant to tidal stripping and heating. 
    The NIHAO emulator prediction of the median radial distribution agrees well with the observations out to ${\sim}150$ kpc.}
    \label{fig:statistics_massive}
\end{figure*}

In \SatGen, the effect of different sub-grid baryonic physics is captured by the halo response relations (\se{initial}).
Among high-resolution cosmological simulations, NIHAO and FIRE feature bursty star formation histories and thus strong, episodic supernovae outflows.
This causes DM cusp-to-core transformations for massive dwarfs ($\Mv\sim10^{10.5}\Msun$ or $\Ms/\Mv\sim10^{-3}$). 
Along with core formation, the overall density profile also becomes less concentrated. 
The APOSTLE and Auriga simulations, on the other hand, have relatively smooth and continuous star formation histories and therefore fewer intense episodes of supernovae feedback.
The DM haloes remain cuspy throughout the mass range simulated \citep{Bose19}.
Cuspy, concentrated systems, once becoming satellites, are more resistant to tidal stripping.
This is taken into consideration by the tidal evolution tracks described in \se{tracks}.

Therefore, as we can anticipate, an APOSTLE-like halo response would yield higher satellite counts than the more bursty NIHAO model. 
This is clearly shown by \figs{statistics_all} and \ref{fig:statistics_massive}.
We note that this effect is more pronounced for massive satellites (as in \fig{statistics_massive}) than for the entire surviving population, which is dominated by low-mass systems (as in \fig{statistics_all}).
Specifically, the NIHAO emulator produces 20\% fewer massive satellites than the APOSTLE emulator, while the difference in the abundance of all surviving satellites ($m>10^6\Msun$) is only $\sim 7\%$.
This is largely due to the fact that the two halo response relations mainly differ in the massive-dwarf regime, converging at the low mass end. 

The relative importance of the halo response versus the baryonic disc of the host, in terms of its influence on satellite abundance, also depends on the model selection -- for the whole population of surviving satellites, the disc effect is dominant, whereas for the massive dwarf subset, the disc effect is comparable to the halo response effect, both contributing to a ${\sim}20-25\%$ difference. 

\fig{statistics_massive} shows that the halo-to-halo variance is dramatic, especially in the satellite spatial distributions.
This highlights the importance of having a large sample if we hope to distinguish between feedback models.
Hydro-simulation suites that consist of on the order of ten MW/M31 analogues would struggle in revealing the aforementioned differences \citep{Samuel20}. 
Similarly, on the observational side, surveys of more MW/M31 analogues are needed. The SAGA survey \citep{Geha17}, which will contain ${\sim}100$ MW-like systems when completed, will start to be a useful observational benchmark for differentiating feedback models based on the demographics of their satellite galaxies.

\subsection{Effect of the disc potential}
\label{sec:DiscEffect}

\begin{figure*}
	\includegraphics[width=0.45\textwidth]{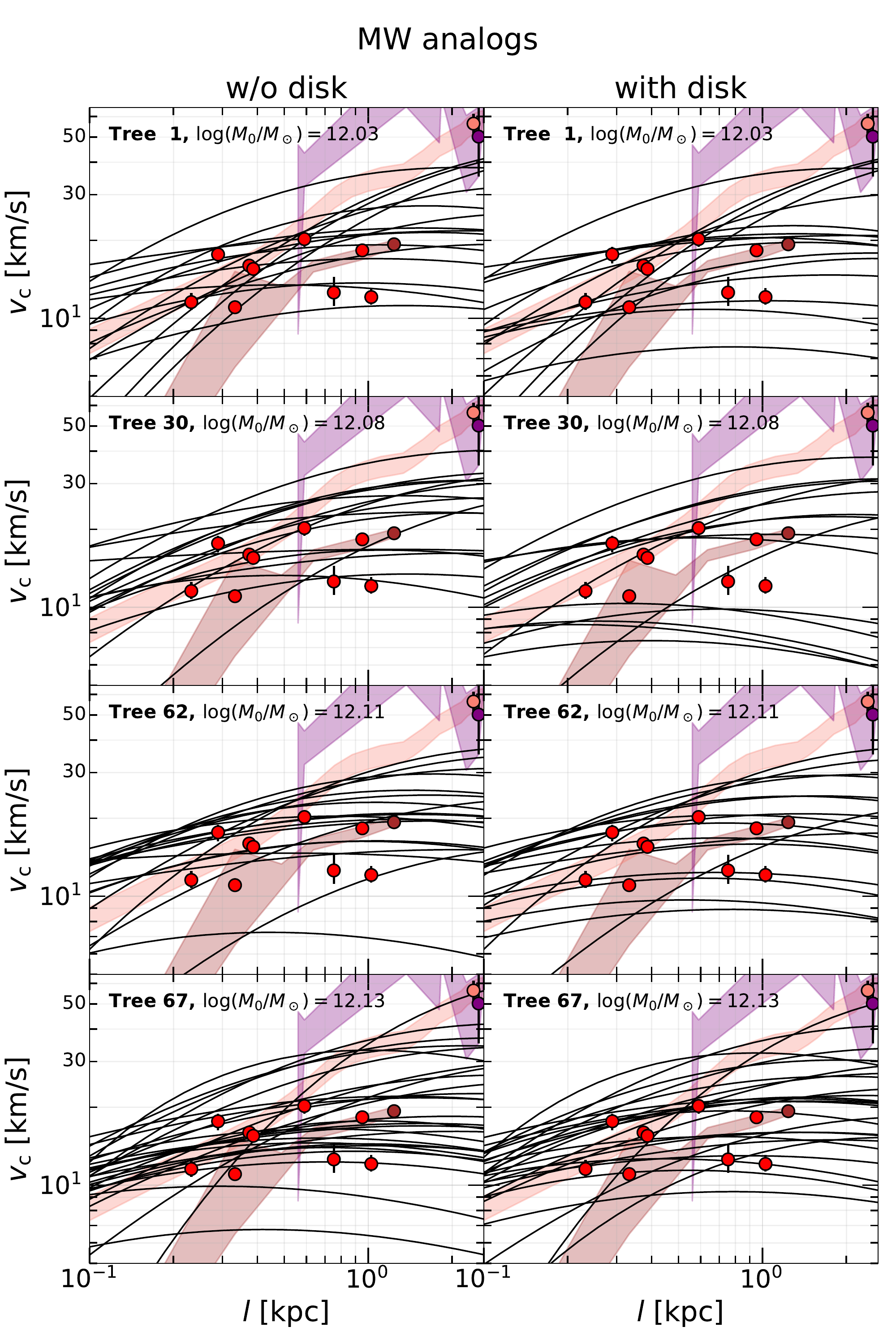}
	\includegraphics[width=0.45\textwidth]{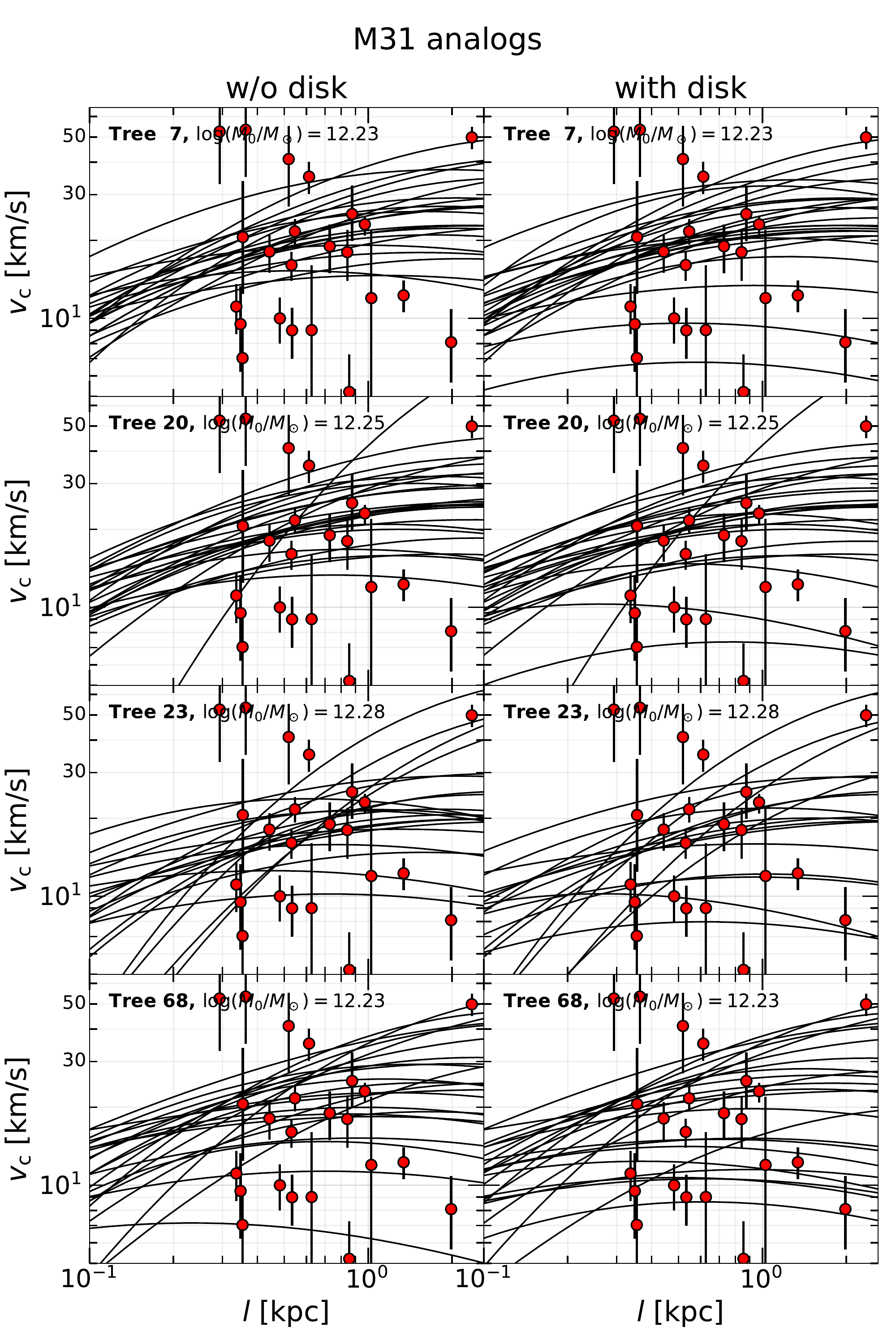}
    \caption{Examples of rotation curves of massive satellites ($\ms>10^5\Msun$) of MW-sized ($\Mv=10^{12-12.15}\Msun$) and M31-sized ($\Mv=10^{12.15-12.3}\Msun$) host haloes at $z=0$, from the NIHAO-emulating models. 
    Each row is a random realization (indicated as ``Tree $i$''), with the left-hand side and right-hand side panels having exactly the same merger history but differing in whether a baryonic disc is included ({\it right}) or ignored ({\it left}) when evolving the satellites (see \se{SHMF} for details about the disc setup). 
    Symbols with error bars are kinematic data from the MW and M31 satellites compiled from the literature, where the red symbols are compiled by \citet{GK19} using the references therein and the brown, pink, and purple symbols and the associated color bands are rotation curves of the Sagittarius dwarf, SMC, and LMC, respectively \citep{Cote00, BS09, MK14}.
    Overall, the model rotation curves are in reasonable agreement with the observed kinematics, especially in the cases with a baryonic disc.
    The disc has a weak but noticeable effect of increasing the diversity of the rotation curves, as can be most clearly seen in {\tt Tree 7}, {\tt Tree 20}, and {\tt Tree 30}.}
    \label{fig:RotationCurves}
\end{figure*}

As we can expect, injecting a baryonic disc into the host galaxy has the effect of depleting satellites.
This is simply because the disc is an extra source of tidal field and dynamical friction in addition to the smooth host halo. 
This satellite-depletion effect has been discussed by, e.g., \citet{Penarrubia10} and \citet{GK17}, using semi-analytical models and simulations. 
Here, we report consistent results.
As shown in the right-hand panel of \fig{statistics_all}, adding a disc reduces the abundance of surviving satellites by ${\sim}20\%$.
This effect is stronger towards the centre of the host and is not very sensitive to the halo response model.

\begin{figure}
	\includegraphics[width=0.47\textwidth]{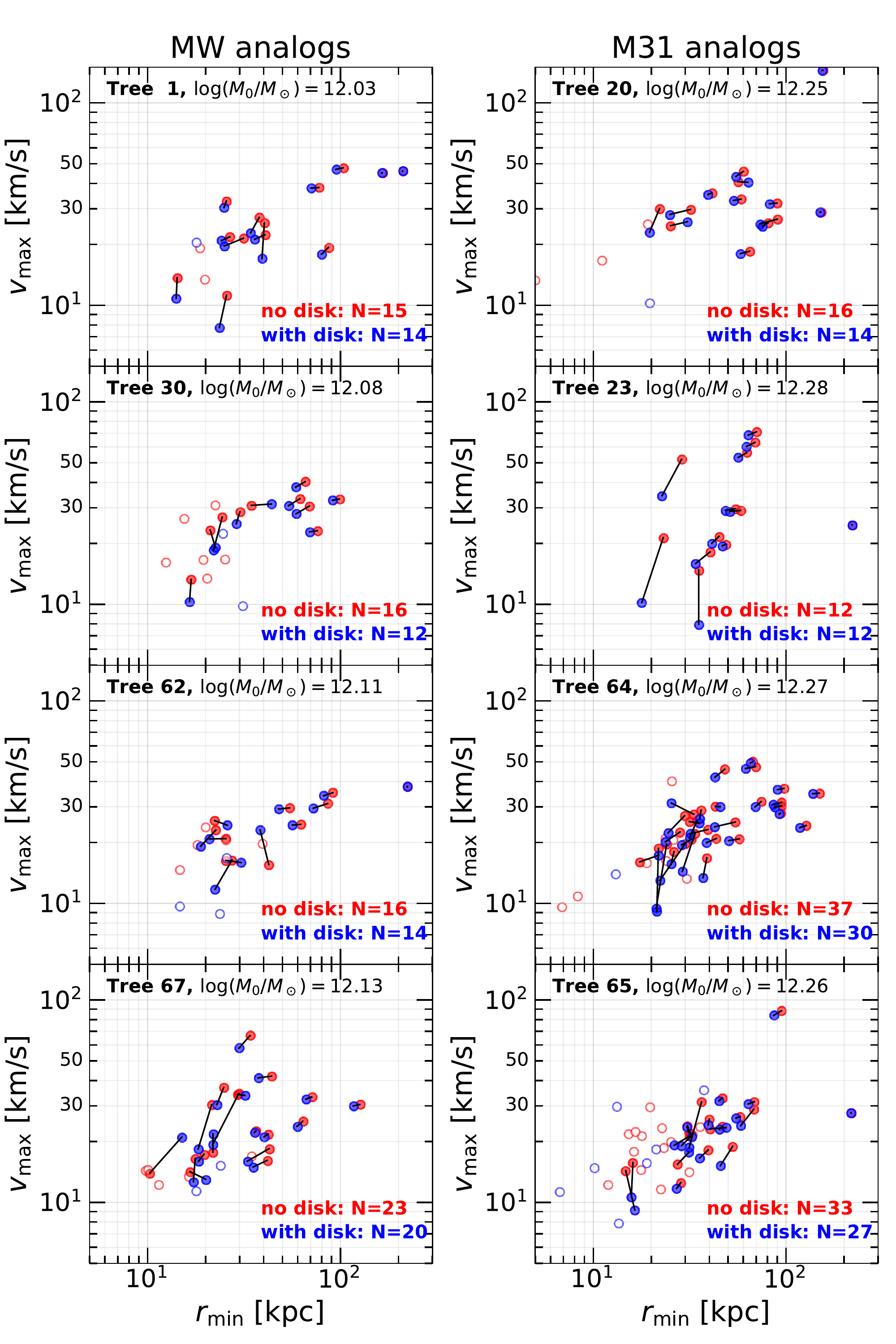}
    \caption{Effect of a disc potential on satellite structure -- $\vmax$ as a function of minimum galactocentric distance, $\rmin$, for surviving massive satellites ($\ms>10^5\Msun$) in MW-sized hosts ($\Mv=10^{12-12.15}\Msun$) and in M31-sized hosts ($\Mv=10^{12.15-12.3}\Msun$). 
    Each panel shows the satellites in a {\it pair} of realizations with an identical, random merger tree evolved with and without a disc. 
    Short black lines connect satellites shared in common (solid symbols) by the models with and without the disc, highlighting the change in $\vmax$. 
    Open symbols represent the massive satellites that only belong to the disc models or the no-disc models. 
    The numbers quoted in the lower right-hand corners of each panel are the numbers of surviving massive satellites. 
    Focusing on the common satellites, we find that the disc generally decreases their $\vmax$ and $\rmin$.
    The $\vmax$ change is more pronounced for those satellites with smaller $\rmin$.}
    \label{fig:StructuralChange}
\end{figure}

In addition to depleting satellites, the disc also plays a secondary role of diversifying satellite structure. 
This is a subtle, but important, effect for reconciling the small-scale issues. 
Notably, the ``too-big-to-fail'' problem (TBTF) can be formulated as a tension between the narrow $\vmax$ distribution of subhaloes from $\Lambda$CDM models and the relatively broad $\vmax$ distribution of the observed massive satellites \citep[e.g.,][]{JB15}.
The cusp-core issue is a tension that arises due to the fact that the observationally inferred DM inner slopes are quite diverse \citep[e.g.,][]{Oman15} whereas the $\Lambda$CDM subhalo inner slopes (in DM-only simulations) are almost exclusively cuspy.  
That is, both the TBTF and the cusp-core issues boil down to a structural diversity issue. 

A commonly used diagnostic for TBTF is the comparison of the rotation curves (RCs) of massive satellites predicted by the model versus the circular velocities at certain radii observed for MW/M31 massive satellites, usually $\vc(\leff)$.
\fig{RotationCurves} presents such examples from our NIHAO-emulating models.
Overall, the agreement between the models and the data is decent, but we focus on comparing the results from the (merger tree-matched) models with and without the disc. 
We can see that the spread of the RCs is marginally larger in the models with a disc. 
This is especially clear in, e.g., {\tt Tree 7}, {\tt Tree 20}, and {\tt Tree 30}. 
In the few cases, such as {\tt Tree 67} and {\tt Tree 68}, where the RCs in the no-disc models appear to be more scattered, the visual impression is actually misled by the fact that there are more satellites in the no-disc model. 
For an abundance-matched comparison, the RCs in the no-disc model are always more narrowly crowded and less diverse.

To better show the disc's role in broadening the structural diversity, we examine in \fig{StructuralChange} the $\vmax$ change as a function of the minimum host-centric distance, $\rmin$, for individual massive ($\ms>10^5\Msun$) satellites in the merger tree-matched models with and without the disc. 
We can see that the disc decreases the $\vmax$ values by up to 50\%, depending on $\rmin$. 
Generally, the closer a satellite gets to the host centre, the more that $\vmax$ decreases with respect to the no-disc case. 
The disc also marginally decreases the minimum galactocentric distances, as can be expected.  

\begin{figure*}
	\includegraphics[width=\textwidth]{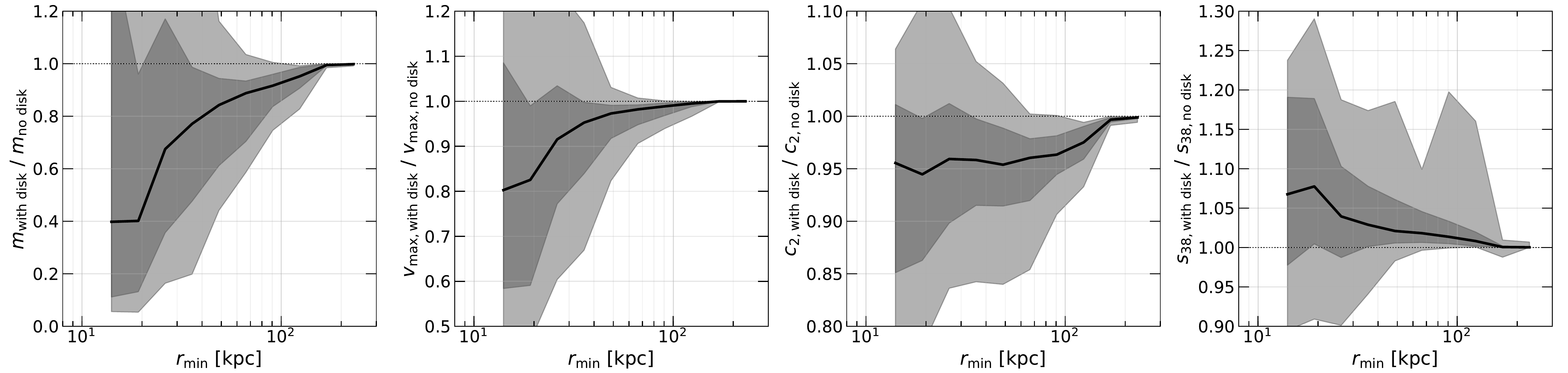}
    \caption{The median ratios of subhalo mass, $\vmax$, concentration, and inner density slope ($s_{38}$, see \se{DiscEffect} for definition) between the models {\it with} and {\it without} the disc potential, all as a function of the minimum galactocentric distance (as measured in models with the disc), for all of the shared massive surviving satellites ($\ms>10^5\Msun$) in all of the 100 random realizations. 
    Darker and lighter shaded bands indicate 16-84 and 3-97 percentiles, respectively. 
    On average, the disc potential decreases satellite mass, $\vmax$, concentration, and increases the density slope -- all in all, the disc increases satellite structural diversity.}
    \label{fig:AverageStructuralChange}
\end{figure*}

\fig{AverageStructuralChange} extends the analysis to the full ensemble, showing the median ratios of subhalo mass ($m_{\rm with\ disc}$/$m_{\rm no\ disk}$), maximum circular velocity ($m_{\rm with\ disc}$/$m_{\rm no disk}$), subhalo concentration ($c_{\rm 2,with\ disc}/c_{\rm 2,no\ disc}$), and logarithmic inner density slope ($s_{\rm 38,with\ disc}/s_{\rm 38,no\ disc}$), as functions of the minimum host-centric distance measured in the simulations with disc, $\rmin$, of massive surviving satellites in all of the 100 realizations.
Here, for the density slope we follow the convention in observational studies to measure it at fixed physical aperture (as opposed to a relative aperture of 0.01$\lv$ that is convenient for theoretical studies) -- in particular, we use the average slope between $l=0.3\kpc$ and $0.8\kpc$, $s_{38}\equiv -\ln[\rho(0.8\kpc)/\rho(0.3\kpc)]/\ln(0.8/0.3)$, following \citet{Relatores19}.
On average, the disc decreases the subhalo mass by up to 60\%, $\vmax$ by 20\%, concentration by 5\%, and steepens the density slope by 8\%. 
Satellites need to reach small galactocentric distances to experience these changes: those not having been within 50 kpc of the galactic centre are barely affected. 

We emphasize again that both the internal and external baryonic effects contribute a ${\sim}25\%$ effect on the abundance and structure of satellite galaxies. 
The halo-to-halo variance due to different merging histories easily overwhelms these baryonic effects, unless large samples are utilized.

\section{Discussion: Survival versus disruption}
\label{sec:discussion}

\begin{figure*}
	\includegraphics[width=0.92\textwidth]{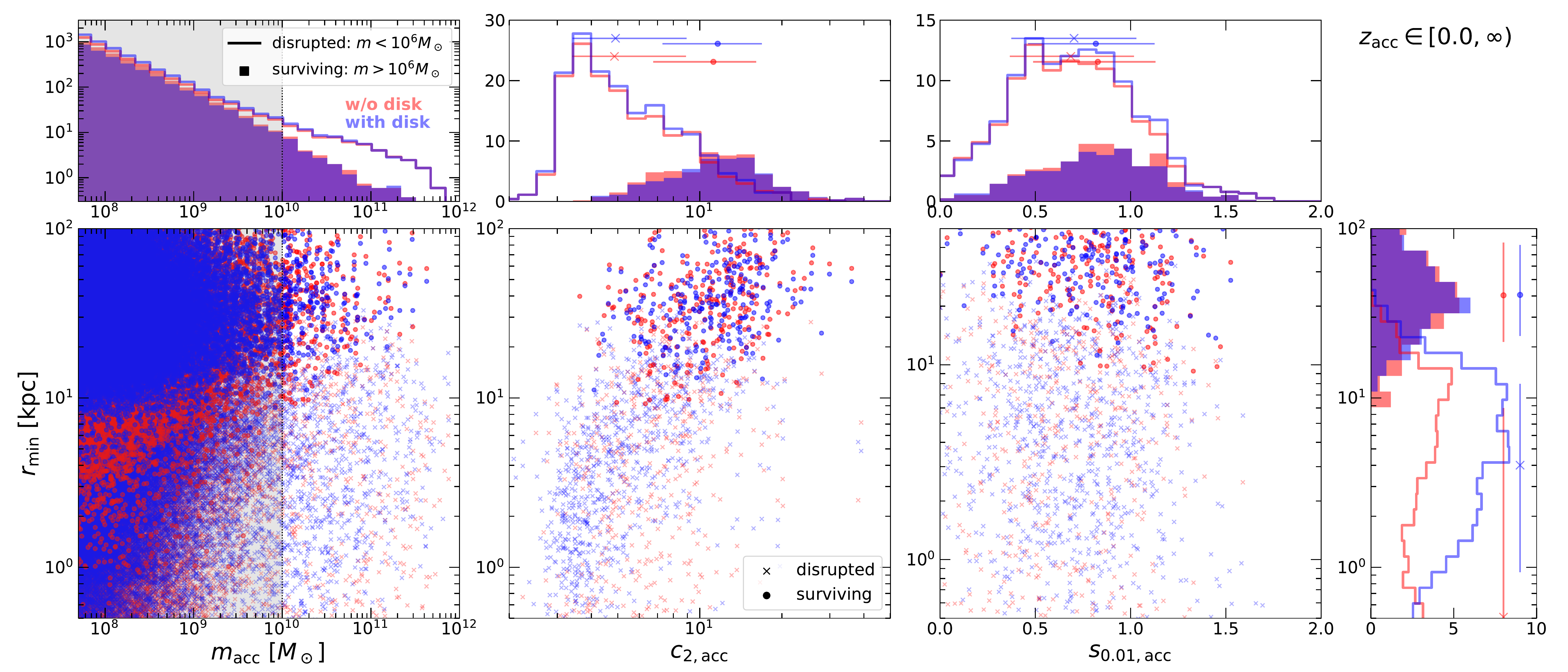}
	\includegraphics[width=0.92\textwidth]{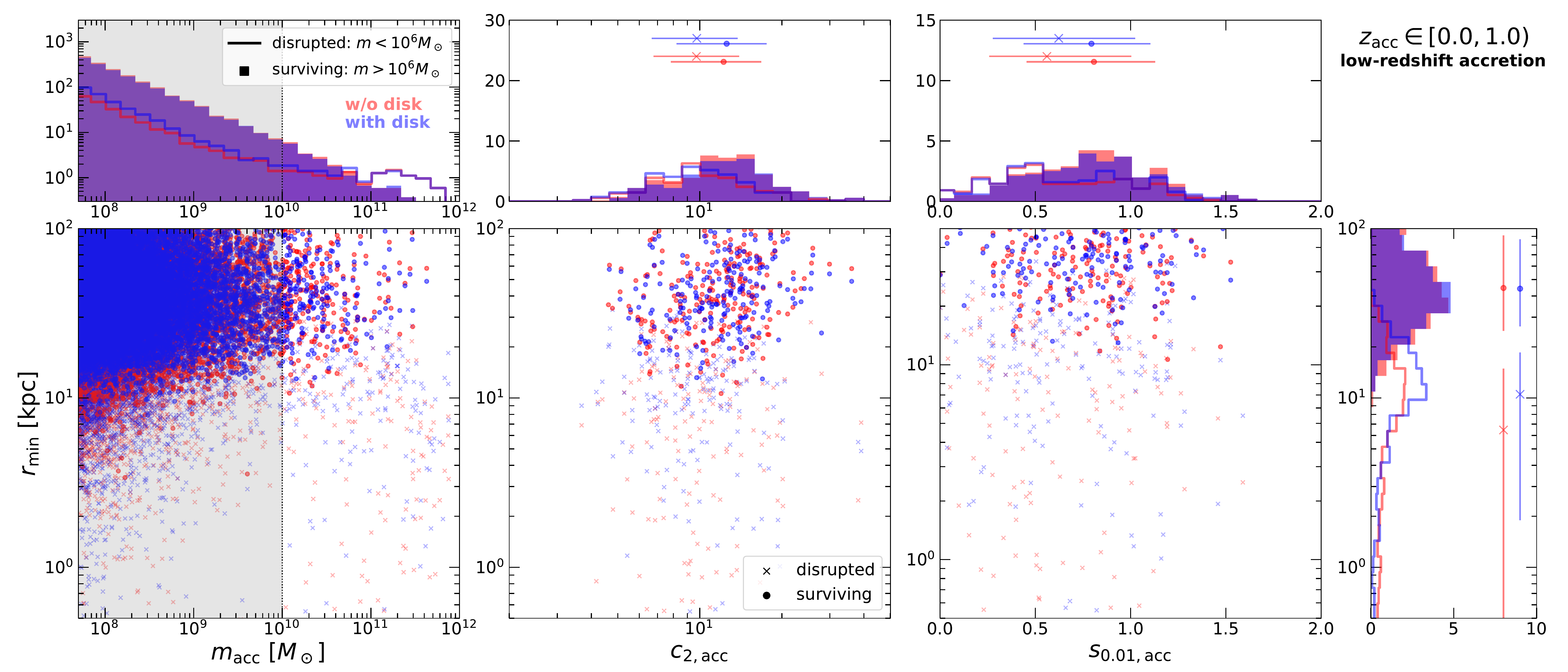}
	\includegraphics[width=0.92\textwidth]{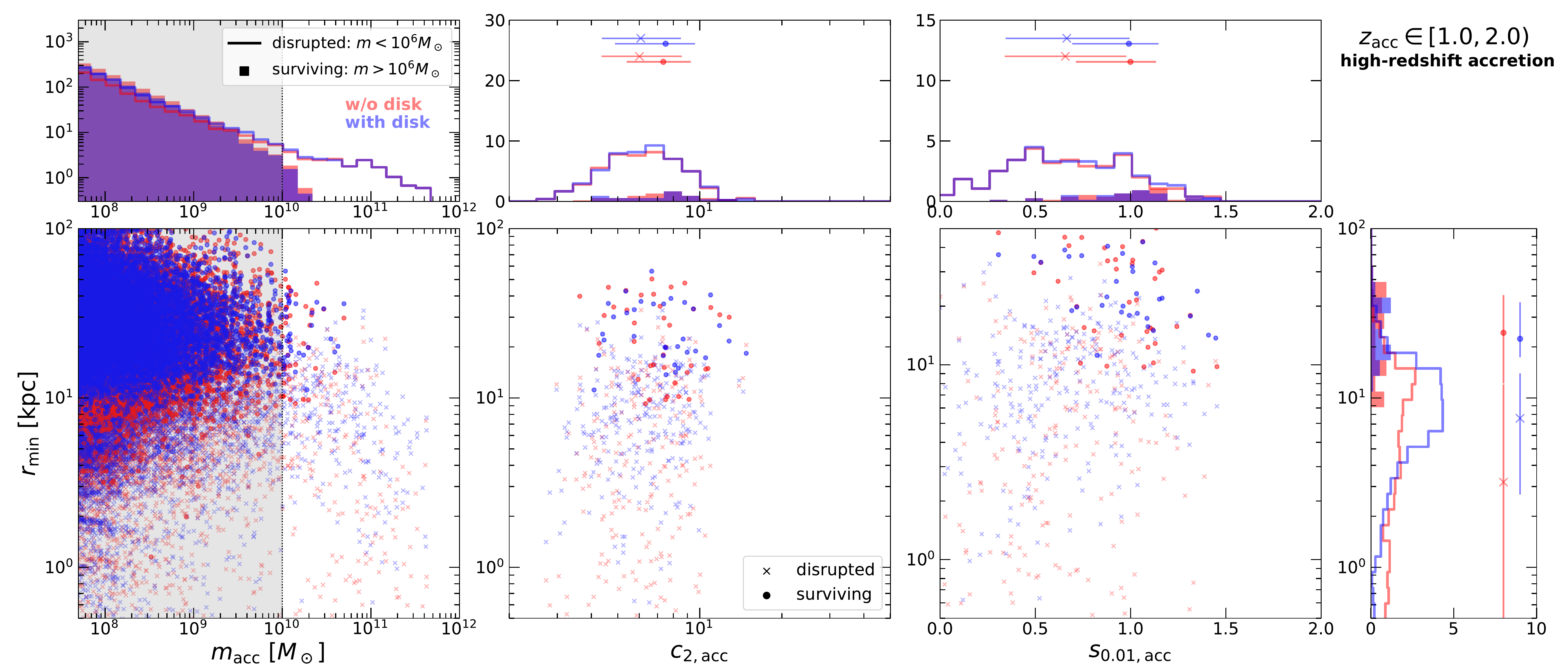}
    \caption{Comparison of disrupted satellites ($m<10^6\Msun$) and surviving satellites ($m>10^6\Msun$) in terms of their minimum host-centric distance versus mass, concentration, and inner slope at accretion, for the NIHAO-emulating models. 
    The first row shows the results for satellites accreted throughout cosmic history. 
    The second and third rows show results for satellites accreted at low redshift ($0\le\za<1$) and higher redshift ($1\le\za<2$), respectively.
    The top and side panels show the 1D marginalized histograms. 
    Surviving satellites are shown as filled histograms while disrupted ones are shown as empty steps.
    The middle column ($\rmin$ versus $c_{\rm 2,acc}$) and right-hand column ($\rmin$ versus $s_{\rm 0.01,acc}$) focus only on satellites with $\ma>10^{10}\Msun$. 
    Key takeaways:
    (1) Disruption occurs throughout the mass range, with a hump at the massive end, illustrating that massive satellites experience stronger dynamical friction.
    (2) Surviving satellites have higher concentration and cuspier density profiles at infall. However, the concentration trend largely reflects a progenitor bias (namely that concentration anti-correlates with redshift) and is significantly reduced if focusing on satellites accreted in the same redshift range. 
    (3) The disc potential causes disruption to occur at larger galacocentric distances.}
    \label{fig:DisruptedVersusSurviving}
\end{figure*}

It is natural to wonder what determines the fate of a satellite --  under what internal and external conditions will a satellite survive, and under what conditions will \sout{a satellite} \new{it} be disrupted? 
With the relatively large statistical samples provided by \SatGen, we can address these questions quantitatively.

\fig{DisruptedVersusSurviving} compares the distributions of surviving satellites ($m>10^6\Msun$) and of disrupted satellites ($m<10^6\Msun$) in the space spanned by the minimum galactocentric distance ($\rmin$) versus virial mass at infall ($\ma$), concentration at infall ($c_{\rm 2,acc}$), and logarithmic inner density slope at infall ($s_{\rm 0.01,acc}$).
In the first row of \fig{DisruptedVersusSurviving}, we include satellites accreted throughout cosmic history, whereas in the second and third rows of \fig{DisruptedVersusSurviving}, we consider satellites accreted at low redshift ($\za<1$) and higher redshift ($\za=1-2$), separately.  
We focus only on the NIHAO emulator results, but compare the models with and without the galactic disc potential. 

There are several features worth mentioning.
First, disruption occurs throughout the infall mass range. 
At the most massive end ($\ma\ga10^{11}\Msun$), disruption actually dominates over survival. 
This can be clearly seen via the $\ma$ distributions of satellites accreted after $z=1$ (the top panel of the second row, first column, of \fig{DisruptedVersusSurviving}).
This massive-end bump highlights the strong satellite mass dependence of dynamical friction: only massive satellites with $m/M\ga0.1$ undergo significant orbital decay. 
We caution that we have arbitrarily defined ``disruption'' as subhalo mass dropping below $10^6\Msun$.
This mass threshold is comparable or slightly better than the mass resolution of state-of-the-art zoom-in simulations of MW-sized haloes, where the DM particle mass is a few times $10^4\Msun$ \citep[e.g.,][]{Wetzel16} and at least 100 particles are needed to resolve a substructure.
Hence, our disruption threshold is comparable to that in high-resolution simulations. 
However, we emphasize that mass dropping below an arbitrary threshold does not necessarily correspond to physical disruption, and we refer interested readers to \citet{vdB18} for a thorough discussion. 

Second, surviving satellites were more concentrated and more cuspy at accretion. 
Specifically, if we focus on massive satellites with $\ma>10^{10}\Msun$, the surviving ones have a median concentration of $c_{\rm 2, acc} \approx11$ and a median inner slope of $s_{\rm 0.01,acc}\approx0.8$, while the disrupted ones have a median concentration of $c_{\rm 2, acc}\approx5$ and a median slope of $s_{\rm 0.01,acc}\approx0.7$. 
At face value, the concentration trend seems to have a simple interpretation: denser haloes are more resistant to tidal disruption.
While this statement is true on its own, it is actually not the main factor at play here.
The time spent in the host halo is more important for the disruption of a subhalo than properties of the initial density profile.  
This can be seen from the second and third rows of \fig{DisruptedVersusSurviving}: selecting satellites by infall redshift significantly reduces the difference in $c_{\rm2,acc}$ between the disrupted and surviving populations. 
Halo concentration at fixed mass anti-correlates with redshift \citep[e.g.,][]{DM14}, so the satellites that were accreted earlier (and thus exposed for a longer time to the tidal field of the host) naturally tend to have lower concentrations. 
However, the inner cuspiness is almost independent of redshift. 
In fact, taking $\za$ bins makes the slope difference more pronounced: for $\za\in[1,2)$, the surviving satellites have $s_{\rm 0.01,acc}\approx1$, and the disrupted ones have $s_{\rm 0.01,acc}\approx0.6$. 

Third, the disc significantly changes the minimum galactocentric distance at which disruption takes place. 
In particular, without a disc potential, satellites can travel to as close as $\rmin\la1\kpc$ from the galactic centre before becoming disrupted, whereas with a disc, most disruption events occur outside 1 kpc, with a median $\rmin$ of 4 kpc. 
This again illustrates the disruptive role of the galactic disc.
Massive surviving satellites can seldom travel within 10 kpc of the galactic centre.
In this way, the Solar neighbourhood is shielded against massive satellites. 

\section{Conclusion}
\label{sec:conclusion}

In this paper, we presented a new semi-analytical model (\SatGen) for generating satellite galaxy populations.
The model is devised to generate statistical samples of satellite galaxy populations for desired host properties, emulating zoom-in cosmological simulations and outperforming simulations in statistical power. 
It combines halo merger trees, empirical relations that describe the galaxy-halo connection, and analytical prescriptions for satellite evolution, incorporating new developments in these areas.
Its improvements and features can be summarized as follows:
\begin{itemize}
\item It uses the \citet{PCH08} algorithm to generate halo merger trees, with parameters recently re-calibrated by \citet{Benson17}. 
It can also be applied to merger trees from $N$-body simulations.
\item It supports halo density profiles that are more flexible than the \citetalias{NFW97} profile, including the \citetalias{Einasto65} profile and the \citetalias{Dekel17} profile, the latter of which has useful analytical properties. 
It also uses the \citetalias{MN75} profile for describing discs.
\item It can be used to emulate hydro-simulations with different sub-grid baryonic physics via an empirical treatment of the halo response to star formation and feedback, as extracted from zoom-in hydro-simulations of field galaxies. 
\item It makes use of stellar-mass-halo-mass relations from halo abundance matching, as well as galaxy-size-halo-size relations extracted from hydro-simulations, in order to initialize the baryonic properties.
\item It supports satellite orbit integration in {\it composite} host potentials, consisting of (combinations of) a DM halo, baryonic disc, and stellar bulge. 
\item It uses tidal evolution tracks obtained from high-resolution idealized simulations from \citet{Penarrubia08, Penarrubia10} and \citet{Errani15, Errani18}, following the structural evolution of satellites.
This, together with the halo response relations, enables \SatGen to propagate the baryonic effects seen in hydro-simulations to the satellite populations -- a task that is difficult for simulations because of the high numerical resolution required. 
\end{itemize}

We presented a proof-of-concept application of \SatGen.
We generated samples much larger than state-of-the-art zoom-in simulations for MW and M31 at comparable numerical resolution.
We experimented with different halo response models, using \SatGen to emulate simulations with bursty star formation and strong feedback (e.g., NIHAO and FIRE) and simulations with smoother star formation, and thus negligible halo response, in massive dwarfs (e.g., APOSTLE and Auriga). 
We also experimented with models with and without a galactic disc potential in order to quantify the influence of the disc on satellite statistics. 
In other words, we explored the internal (halo response) and external (host-disc) baryonic effects on satellite properties.
The conclusions of this study are as follows:
\begin{itemize}
\item We find that the model predictions of the $\vmax$ function, rotation curves, and spatial distributions of bright satellites with $\ms>10^5\Msun$ are in good agreement with observations. 
This is achieved without fine-tuning model parameters. 
\item Different halo response models yield slightly different satellite abundances: on average, the NIHAO emulator yields 25\% less satellites with $\ms>10^5\Msun$ within 300 kpc of the galactic centre than the APOSTLE emulator. The effect is smaller if we include all of the surviving satellites, illustrating the fact that the difference in the halo response is most prominent for massive dwarfs. 
Given the large halo-to-halo variance as revealed by the model, and given the limited observational sample, it currently remains difficult to use the observed satellite spatial distribution to distinguish between the two feedback patterns.
\item Adding a disc potential to the host causes, on average, a 20\% (30\%) reduction in satellite number count within 300 (100) kpc. 
In addition to satellite depletion, the disc slightly increases the structural diversity of massive satellite dwarfs. 
On average, a disc decreases the satellite $\vmax$ by up to 20\%, concentration by up to 5\%, and increase the density slope measured at the fixed physical aperture of $0.3-0.8$ kpc by up to 8\%, depending on the minimum galactocentric distance that the satellite can reach. 
This helps with alleviating the small-scale problems of $\Lambda$CDM. 
\item The fate of a massive satellite galaxy ($\ma>10^{10}\Msun$) depends on how close it gets to the galactic centre: the surviving satellites seldom reach within 10 kpc of the centre, whereas the disrupted ones have a minimum galactocentric distance of $\rmin{\sim}4$ kpc (or $\la1$ kpc if there was no galactic disc). 
The fate also depends on the initial structure at infall: more concentrated and cuspier haloes are more likely to survive. 
However, the concentration trend is largely due to a progenitor bias, in the sense that satellites that have been exposed to the tidal field for a longer time, i.e., those that were accreted earlier, have lower concentration at accretion because of the anti-correlation between halo concentration and redshift. 
\end{itemize}

Overall, we have shown that \SatGen can emulate numerical simulations of very high resolution decently, capturing the bulk of the baryonic effects on the abundance, spatial distribution, and internal structure of satellites. 
Thanks to the tidal evolution recipes that are extracted from high-resolution idealized simulations, it avoids the numerical artifacts of over-stripping. 
Simulating a statistically large sample of MW/M31-sized systems, not to mention galaxy groups or clusters, while retaining the resolution for satellite dwarfs is computationally challenging for numerical simulations.
Therefore, the \SatGen model complements simulations nicely in terms of statistical power.
In an upcoming work (Jiang et al., in prep), we use \SatGen to study satellites of group-sized hosts and explore the conditions for forming ultra-diffuse galaxies and compact dwarf satellites.
The \SatGen code is made publicly available at \href{https://github.com/shergreen/SatGen}{https://github.com/shergreen/SatGen}.


\section*{Acknowledgements}

The authors are thankful to Yuval Birnboim, Timothy Carleton, Nicolas Cournuault, Andrew Emerick, Omri Ginzburg, Sharon Lapiner, Mariangela Lisanti, Lina Necib, Jacob Shen, Oren Slone, and Coral Wheeler for helpful discussions.
FJ is supported by the Israeli Planning and Budgeting Committee (PBC) Fellowship, and by the Troesh Fellowship from the California Institute of Technology. 
FJ is thankful to Jo Bovy for publicly sharing his code design wisdom through the software {\tt galpy}. 
FCvdB is supported by the National Aeronautics and Space Administration through Grant Nos. 17-ATP17-0028 and 19-ATP19-0059 issued as part of the Astrophysics Theory Program, and received addition support from the Klaus Tschira foundation.
SBG is supported by the US National Science Foundation Graduate Research Fellowship under Grant No. DGE-1752134.




\bibliographystyle{mnras}
\bibliography{SatGen} 



\appendix

\section{Analytics of profiles}
\label{app:profiles}

Here, we provide the analytical expressions for the profiles of density ($\rho$), enclosed mass ($M$), gravitational potential ($\Phi$), the $R$-component and $z$-component of gravitational acceleration in the cylindrical coordinate system ($f_R$, $f_z$), and the one-dimensional velocity dispersion for an isotropic velocity distribution ($\sigma$), as well as a few convenient relations among the parameters, for each of the potential well classes supported in \SatGen.

\subsection{\citetalias{NFW97}}

We specify an \citetalias{NFW97} profile using the virial mass, $\Mv$, the concentration parameter, $c_2$ (or the corresponding scale radius $\rs=\rv/c_2$), and the average spherical overdensity, $\Delta$. 

\be
\label{eq:NFWdensity}
\rho(r) = \frac{\rho_0}{x\left(1+x\right)^2},\,\,\, {\rm where} \,\,\, x=
\frac{r}{\rs} \,\,\,  {\rm and} \,\,\,  \rho_0 = \frac{c_2^3}{3f(c_2)}\Delta \rhoc, 
\ee
with $ f(x)=\ln(1+x) - x/(1+x)$.

\be
\label{eq:NFWmass}
M(r) = \Mv\frac{f(x)}{f(c_2)}.
\ee

\be
\label{eq:NFWpotential}
\Phi(r) = \Phi_0 \frac{\ln(1+x)}{x} ,\,\,\, {\rm where} \,\,\, \Phi_0 = -4 \pi G \rho_0 \rs^2.
\ee

\bad
f_R = -\frac{\partial\Phi}{\partial R} = \Phi_0 \frac{f(x)}{x} \frac{R}{r^2}\quad {\rm and} \quad
f_z = -\frac{\partial\Phi}{\partial z} = \Phi_0 \frac{f(x)}{x} \frac{z}{r^2},
\ead
where $r = \sqrt{R^2+z^2}$.

\bad
\label{eq:NFWpotential}
\sigma^2(r) &= \Vv^2 \frac{c}{f(c)}x(1+x)^2\int_{x}^{\infty} \frac{f(\xp)}{\xp{}^3(1+\xp)^2}  \rmd \xp\\
&\approx  \Vmax^2 \left(\frac{1.4393 x^{0.354}}{1+1.1756x^{0.725}}\right)^2,
\ead
where the second line is an approximation accurate to $1\%$ for $x=0.01$-$100$ \citep[][see also an analytical solution involving non-elementary functions by \citealt{LM01}]{ZB03}.

The location of the peak circular velocity, $\rmax$, is related to the scale radius, $\rs$, by
\be
\rmax \approx 2.163\rs,
\ee
where $\rs$ is the location at which the logarithmic density slope is 2, $r_2$. 

\subsection{\citetalias{Dekel17}}

We specify a \citetalias{Dekel17} profile using the virial mass, $\Mv$, a concentration parameter, $c$ (or the corresponding scale radius $\rs=\rv/c$), the innermost logarithmic density slope, $\alpha \equiv -\rmd\ln\rho/\ln r|_{r\to 0}$, and the average spherical overdensity, $\Delta$.

\bad
\rho(r) &= \frac{\rho_0}{x^\alpha (1+x^{1/2})^{2(3.5-\alpha)}}, \\
 & {\rm where} \,\,\, x=
\frac{r}{\rs} \,\,\,  {\rm and} \,\,\,  \rho_0 = \frac{c^3(3-\alpha)}{3f(c,\alpha)}\Delta \rhoc, 
\ead
with $f(x,\alpha) =  \chi^{2(3-\alpha)}$ and $\chi = x^{1/2}/(1+x^{1/2})$.

\be 
M(r) = \Mv\frac{f(x,\alpha)}{f(c,\alpha)}.
\ee

\be 
\Phi(r) = -\Vv^2 \frac{2c}{f(c,\alpha)}\left[\frac{1-\chi^{2(2-\alpha)}}{2(2-\alpha)} - \frac{1-\chi^{2(2-\alpha)+1}}{2(2-\alpha)+1} \right].
\ee

\bad
f_R (R,z)&= (2-\alpha)[2(2-\alpha)+1]  \Phi_0  \frac{f(x,\alpha)}{x} \frac{R}{r^2} \quad {\rm and} \\
f_z (R,z)&= (2-\alpha)[2(2-\alpha)+1]  \Phi_0  \frac{f(x,\alpha)}{x} \frac{z}{r^2},\\
& {\rm where}\,\,\, \Phi_0= - \frac{4\pi G \rho_0 \rs^2 }{(3-\alpha)(2-\alpha)[2(2-\alpha)+1]} .
\ead

\bad
\sigma^2(r) &=  \Vv^2 \frac{c}{f(c,\alpha)} \frac{x^{3.5}}{\chi^{2(3.5-\alpha)}} \int_{x}^{\infty}\frac{\chi(\xp)^{4(3-\alpha)+1}}{ \xp{}^{5.5}}\rmd\xp \\
&= 2\Vv^2 \frac{c}{f(c,\alpha)} \frac{x^{3.5}}{\chi^{2(3.5-\alpha)}} \sum_{i=0}^{8} \frac{(-1)^i 8!}{i!(8-i)!}  \frac{1-\chi^{4(1-\alpha)+i}}{4(1-\alpha)+i}.
\ead

We refer interested readers to \citet{Freundlich20} for the analytical expressions of the \citetalias{Dekel17} profile for gravitational lensing-related quantities, including the surface density, deflection angle, shear, and magnification. 

Unlike \citetalias{NFW97}, for which $\rs=r_2$, the \citetalias{Dekel17} scale radius is related to $r_2$ by
\be
r_2 = r_s\left(\frac{2-\alpha}{1.5}\right)^2,
\ee 
such that the relation between the \citetalias{Dekel17} concentration ($c$) and the conventional concentration ($c_2$) is
\be
c_2 = \frac{\rv}{r_2} = \left(\frac{1.5}{2-\alpha}\right)^2 c.
\ee
The locatin of peak circular velocity, $\rmax$, is related to $r_2$ by 
\be
\rmax = 2.25r_2 = (2-\alpha)^2 \rs.
\ee
The profile of the logarithmic density slope is 
\be
s(r) = - \frac{\rmd\ln\rho}{\rmd\ln r} = \frac{\alpha+3.5\sqrt{x}}{1+\sqrt{x}}.
\ee
The slope at $0.01\rv$ is
\be
s_{0.01} \equiv s(0.01\rv) = \frac{\alpha+0.35\sqrt{c}}{1+0.1\sqrt{c}}.
\ee
For $s_{0.01}$ values that are commonly seen in simulations and observations ($0-2$) and for a typical concentration (e.g., $c=10$), we have $\alpha\in(-1.11,1.53)$.
That is, $\alpha$ can be negative for realistic profiles, and thus $s_{0.01}$ is a more physically meaningful quantity than $\alpha$ when it comes to comparing the cuspiness of density profiles. 

\subsection{\citetalias{Einasto65}}

We define an \citetalias{Einasto65} profile using the virial mass, $\Mv$, the concentration parameter, $c$ (or the corresponding scale radius $\rs=\rv/c_2$), the shape index, $n$, and the average spherical overdensity, $\Delta$. 

\bad
\rho(r) &= \rho_0 e^{-x(r)},\\ 
&{\rm where}\,\,\, x = 2n\left(\frac{r}{\rs}\right)^{\frac{1}{n}}\,\,\, {\rm and}\,\,\, \rho_0 = \frac{\Mv}{4\pi h^3  n \gamma[3n,x(\rv)]},
\ead
with $h = \rs/(2n)^n$ and $\gamma(a,x)$ is the non-normalized lower incomplete gamma function.
Here, we have adopted the notations in \citet{RM12} for compact expressions.  

\be
M(r) = M_{\rm tot} \tilde{\gamma}(3n,x),\,\,\,{\rm with}\,\,\,M_{\rm tot} = 4\pi\rho_0 h^3 n \Gamma(3n),
\ee
where $\Gamma(a)$ and $\tilde{\gamma}(a,x)=\gamma(a,x)/\Gamma(a)$ are the Gamma function and the normalized lower incomplete gamma function, respectively.

\be
\Phi(r) = -\frac{GM_{\rm tot}}{h}\left[ \frac{\tilde{\gamma}(3n,x)}{x^n} + \frac{\Gamma(2n,x)}{\Gamma(3n)}\right],
\ee
where $\Gamma(a,x)$ is the non-normalized upper incomplete gamma function.

\bad
f_R (R,z)&=  - G M_{\rm tot} \tilde{\gamma}(3n,x) \frac{R}{r^3}\quad {\rm and} \\
f_z (R,z)&= - G M_{\rm tot} \tilde{\gamma}(3n,x) \frac{z}{r^3}.
\ead

\be
\sigma^2(r) = \frac{GM_{\rm tot}}{h} n e^x \int_x^\infty \frac{ \tilde{\gamma}(3n,\xp) }{e^{\xp} \xp{}^(n+1)} \rmd \xp.
\ee

Like the \citetalias{NFW97} profile, the \citetalias{Einasto65} scale radius, $\rs$, is the same as $r_2$, where the logarithmic density slope is 2. 
The radius of peak circular velocity is related to $\rs$ by
\be
\rmax \approx 1.715\alpha^{-0.00183}(\alpha+0.0817)^{-0.179488} \rs
\ee
\citep{GK14b}.
The profile of the logarithmic density slope is
\be
s(r) =  - \frac{\rmd\ln\rho}{\rmd\ln r} = \frac{x(r)}{n},
\ee
so
\be
s_{0.01} = 2(0.01c_2)^{\frac{1}{n}}.
\ee

\subsection{\citetalias{MN75}}

We define a \citetalias{MN75} profile using the disc mass, $\Md$, a scale radius, $a$, and a scale height, $b$.

\be
\rho(R,z) = \frac{\Md b^2}{4\pi}\frac{aR^2+(a+3\zeta)(a+\zeta)^2}{\zeta^3[R^2+(a+\zeta)^2]^{5/2}},
\ee
where $\zeta=\sqrt{z^2+b^2}$.

\be
M(r) = \frac{\Md r^3}{[r^2+(a+b)^2]^{1.5}},\,\,\,{\rm where}\,\,\,r=\sqrt{R^2+z^2}.
\ee

\be
\Phi(R,z) = -\frac{G\Md}{\sqrt{R^2 + (a+\zeta)^2}}.
\ee

\bad
f_R (R,z)&= - \frac{G\Md}{[R^2+(a+\zeta)^2]^{1.5}} R\quad {\rm and}\\
f_z (R,z)&= - \frac{G\Md}{[R^2+(a+\zeta)^2]^{1.5}} \frac{a+\zeta}{\zeta} z.
\ead

\be
\sigma^2(R,z) =  \frac{G\Md^2 b^2}{8\pi\rho(R,z)}\frac{(a+\zeta)^2}{\zeta^2[R^2+(a+\zeta)^2]^3}.
\ee

The relation between half-mass radius, $\reff$, and the scale lengths, $(a,b)$, is
\be
\label{eq:MNsize}
a = \frac{0.766421}{1+b/a} \reff.
\ee


\section{Structure of evolved satellites}
\label{app:tracks}

\subsection{Tidal evolution tracks}

We use the tidal evolution tracks of \citet{Penarrubia10} for determining the profiles of evolved subhaloes and those of \citet{Errani18} for updating the stellar masses and half-stellar-mass radii.  
These tidal tracks can be expressed with the universal functional form of
\be
g(x) = \left(\frac{1+\xs}{x+\xs}\right)^\mu x^\eta,
\ee 
where, for the DM subhalo, $g$ represents $\vmax(t)/\vmax(0)$ or $\lmax(t)/\lmax(0)$, and $x$ stands for the bound mass fraction $m(t)/m(0)$.
For the stellar component, $g$ represents $\ms(t)/\ms(0)$ or $\leff(t)/\leff(0)$, and $x$ stands for $\mmax(t)/\mmax(0)$, with $\mmax=m(\lmax)$. 
The parameters $\mu$ and $\eta$ depend on the initial logarithmic density slope, $s_{0.01}$ ($\equiv-\rmd\ln\rho/\rmd\ln r|_{r=0.01\rv}$), and $\xs$ depends on the initial stellar size with respect to the initial radius of peak circular velocity of the hosting subhalo, $\leff(0)/\lmax(0)$.
\citet{Penarrubia10} and \citet{Errani18} obtained best-fit parameters for different initial structures ($s_{0.01}=0,0.5,1,1.5$ and $\leff(0)/\lmax(0)=0.05,0.1$) by calibrating the model against idealized $N$-body simulations, which we summarize here in \tab{tracks}.
For the initial structures not listed in the table but within the range of the tabulated initial structures, we use cubic spline interpolation to get the parameters. 
For the initial structures beyond the tabulated range, we do not extrapolate, but use the nearest neighbours in the table. 

\begin{figure*}
	\includegraphics[width=0.75\textwidth]{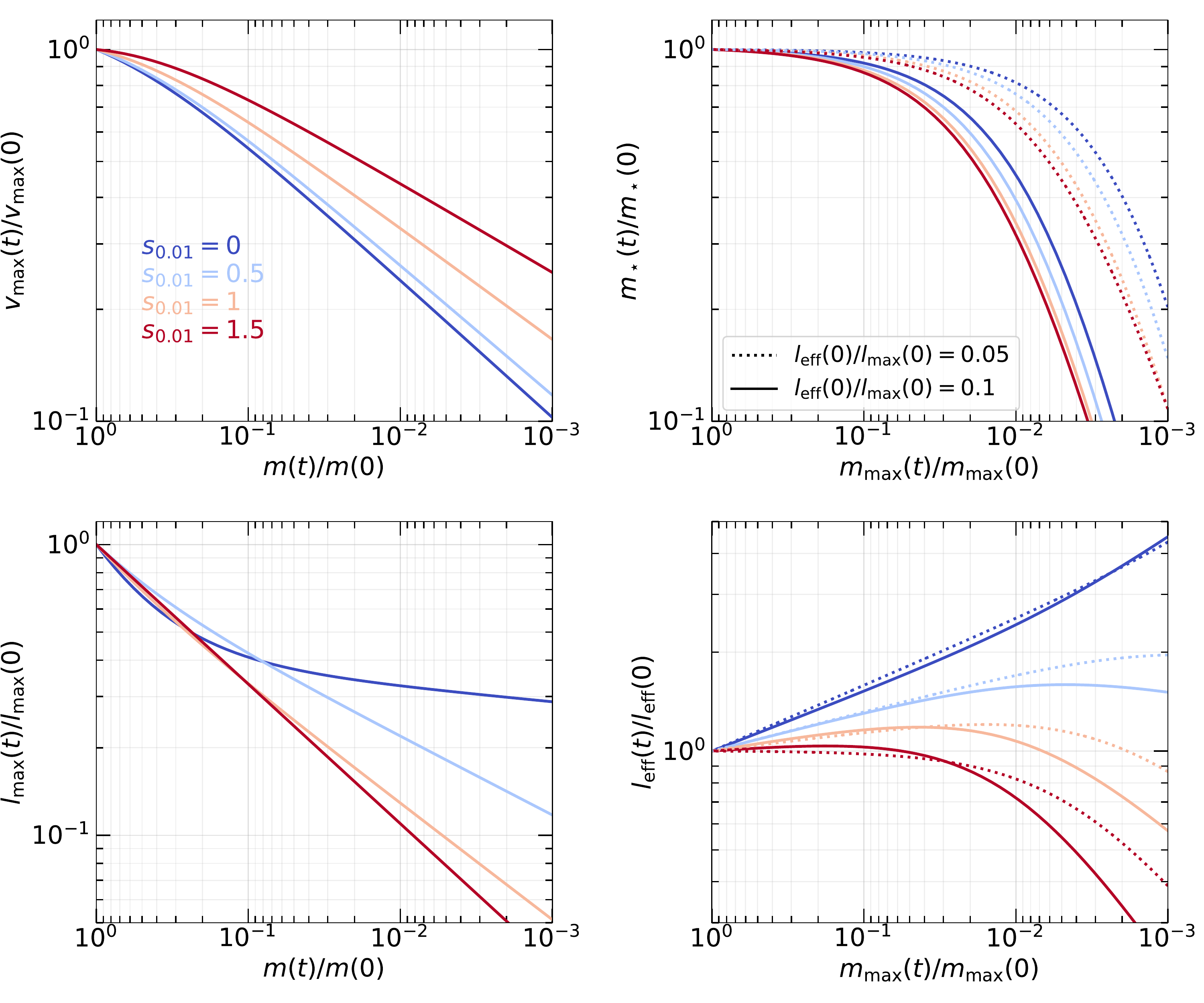}
    \caption{Tidal evolution tracks, compiled from \citet{Penarrubia10} and \citet{Errani18} -- instantaneous subhalo $\vmax$ and $\lmax$ in units of their initial values, both as functions of the instantaneous bound mass fraction, $m(t)/m(0)$ ({\it left}); instantaneous stellar mass, $m$, and half-stellar-mass radius, $\leff$, in units of their initial values, both as functions of the instantaneous ratio between the subhalo mass within $\lmax$ (i.e., $\mmax\equiv m(\lmax)$) and the initial value of $\mmax$. The tracks depend on the initial inner density slope ($s_{0.01}$), and for the stellar component, also depend on the initial compactness of the stellar distribution (as parameterized by $\leff(0)/\lmax(0)$).}
    \label{fig:test_TidalTracks}
\end{figure*}

\fig{test_TidalTracks} illustrates these tidal tracks. 
Note that stellar mass loss is marginal when the subhalo mass within $\lmax$ decreases by $\la90\%$, especially when the initial stellar mass distribution is compact (e.g., when $\leff(0)/\lmax(0)=0.05$).
Also note that, generally, satellite size increases with subhalo mass loss, which manifests due to tidal heating and re-virialization in response to tidal stripping and heating. 
Only cuspy satellites ($\alpha\ga1$) become more compact in stellar size, and the size decrease occurs only after significant subhalo mass loss, when $\mmax(\lmax)$ decreases by $\ga 99\%$. 
This is, however, a viable channel for making compact bright dwarfs ($\ms\sim10^{7-9}$ and $\leff\la 1\kpc$) from massive cuspy galaxies.

\begin{table*}
	\centering
	\caption{Tidal-evolution tracks of the functional form $g(x)=[(1+\xs)/(x+\xs)]^\mu x^\eta$, compiled from \citet{Penarrubia10} and \citet{Errani18} -- for subhaloes, $g$ represents $\vmax(t)/\vmax(0)$ or $\lmax(t)/\lmax(0)$ and $x$ stands for the bound mass fraction, $m(t)/m(0)$; for stellar components, $g$ represents $\ms(t)/\ms(0)$ or $\leff(t)/\leff(0)$, and $x$ stands for $\mmax(t)/\mmax(0)$, where $\mmax=m(\lmax)$. The parameter values in brackets are from linear interpolation/extrapolation.}
	\label{tab:tracks}
	\begin{tabular}{@{\extracolsep{2pt}} | cc | ccc | ccc | ccc | ccc | @{}}
	\toprule
	\multicolumn{2}{ c }{  \backslashbox{initial structure}{quantity} }  & \multicolumn{3}{c}{$g(x)=\frac{\vmax(t)}{\vmax(0)},x=\frac{m(t)}{m(0)}$}  & \multicolumn{3}{c}{$g(x)=\frac{\lmax(t)}{\lmax(0)},x=\frac{m(t)}{m(0)}$} & \multicolumn{3}{c}{$g(x)=\frac{\leff(t)}{\leff(0)}, x=\frac{\mmax(t)}{\mmax(0)}$} & \multicolumn{3}{c}{$g(x)=\frac{\ms(t)}{\ms(0)},x=\frac{\mmax(t)}{\mmax(0)}$} \\
	\cmidrule(l){1-2} \cmidrule(l){3-5} \cmidrule(l){6-8} \cmidrule(l){9-11} \cmidrule(l){12-14} 
	$s_{0.01}$ & $\frac{\leff(0)}{\lmax(0)}$ &  $\mu$ & $\eta$ & $\log\xs$ & $\mu$ & $\eta$ & $\log\xs$ & $\mu$ & $\eta$ & $\log\xs$ & $\mu$ & $\eta$ & $\log\xs$  \\
	\cmidrule(l){1-2} \cmidrule(l){3-5} \cmidrule(l){6-8} \cmidrule(l){9-11} \cmidrule(l){12-14}
	\multirow{2}{*}{1.5} & 0.05 & \multirow{2}{*}{0.4}& \multirow{2}{*}{0.24} & \multirow{2}{*}{0} & \multirow{2}{*}{0} & \multirow{2}{*}{0.48} & \multirow{2}{*}{0} & (0.59) & (0.59) & (-2.4) & (1.39) & (1.39) & (-2.4)  \\
	\cmidrule(l){2-2}   \cmidrule(l){9-11} \cmidrule(l){12-14}
	                              & 0.1 &  &  &  &  &  &  & (0.75) & (0.71) & (-2.0) & (1.68) & (1.68) & (-2.0)   \\
	\cmidrule(l){1-2} \cmidrule(l){3-5} \cmidrule(l){6-8} \cmidrule(l){9-11} \cmidrule(l){12-14}
	\multirow{2}{*}{1}  & 0.05 & \multirow{2}{*}{0.4} & \multirow{2}{*}{0.3} & \multirow{2}{*}{0} & \multirow{2}{*}{-0.3} & \multirow{2}{*}{0.4} & \multirow{2}{*}{0} & 0.47 & 0.41 & -2.64 & 1.87 & 1.87 & -2.64   \\
	\cmidrule(l){2-2}   \cmidrule(l){9-11} \cmidrule(l){12-14}
	                             & 0.1 &  &  &  & &  &  & 0.5 & 0.42 & -2.08 & 1.8 & 1.8 &  -2.08 \\
	\cmidrule(l){1-2} \cmidrule(l){3-5} \cmidrule(l){6-8} \cmidrule(l){9-11} \cmidrule(l){12-14}
	\multirow{2}{*}{0.5} & 0.05 & \multirow{2}{*}{0.4} & \multirow{2}{*}{0.35} & \multirow{2}{*}{0} & \multirow{2}{*}{-0.4}  & \multirow{2}{*}{0.27} & \multirow{2}{*}{0} & (0.19) & (0.07) & (-2.9) & (2.35) & (2.35) & (-2.9) \\
	\cmidrule(l){2-2}   \cmidrule(l){9-11} \cmidrule(l){12-14}
	                               & 0.1 &  &  &  &   &  &  & (0.21) & (0.09) & (-2.2) & (1.93) & (1.93) & (-2.2)  \\
	\cmidrule(l){1-2} \cmidrule(l){3-5} \cmidrule(l){6-8} \cmidrule(l){9-11} \cmidrule(l){12-14}
	\multirow{2}{*}{0}  & 0.05 &  \multirow{2}{*}{0.4} & \multirow{2}{*}{0.37} & \multirow{2}{*}{0} & \multirow{2}{*}{-1.3} & \multirow{2}{*}{0.05} & \multirow{2}{*}{0} & -0.15 & -0.35 & -3.12 & 2.83 & 2.83 & -3.12  \\
	\cmidrule(l){2-2}   \cmidrule(l){9-11} \cmidrule(l){12-14}
	                             & 0.1 &  &  &  &   &  &  & -0.15 & -0.33 & -2.33 & 2.05 & 2.05 & -2.33  \\
	\bottomrule
	\end{tabular}
\end{table*}

\subsection{Evolved subhalo profiles}

The parameters that we use to define the subhalo profiles -- e.g., for the \citetalias{Dekel17} profile -- $c$, $\alpha$, and $\Delta$, are not directly provided by the tidal tracks. 
We need to translate $(\vmax,\lmax)$ to $(c,\alpha,\Delta)$ in order to update the profiles of evolved subhaloes.
Note that the evolved subhaloes have higher overdensities ($\Delta$) compared to distinct haloes, which have $\Delta=200$.

Since the number of parameters $(c,\alpha,\Delta)$ exceeds that of the constraints $(\vmax,\lmax)$, we need an additional assumption. 
We follow \citet{Penarrubia10} to assume that the innermost slope $\alpha$ is constant.
One can analytically show that the innermost part of a subhalo is adiabatically shielded against tidal shocks \citep{GHO99}.
In addition, several numerical studies have shown that the logarithmic density slope at $l\to 0$ barely changes even if the subhalo is stripped down to $0.1\%$ of its initial mass \citep{Penarrubia10, vdB18, BO18}.
Under this assumption, we can express $c$ and $\Delta$ in terms of $\vmax$ and $\lmax$. 
We use two relations,
$\rmd \vc^2/\rmd l |_{\lmax} = 0$ and $\vmax^2 = \vc^2(\lmax)$,
which give
\be
\label{eq:EvolvedConcentration}
c = (2-\alpha)^2 \frac{\lv}{\lmax}
\ee
and
\be
\vmax^2 = \frac{G\mv}{\lmax} \frac{f[(2-\alpha)^2,\alpha]}{f(c,\alpha)},
\ee
where $f(x,\alpha) =  \chi^{2(3-\alpha)}$ and $\chi = x^{1/2}/(1+x^{1/2})$.
Combining these two relations, we can express the evolved virial mass ($\lv$) and thus the evolved overdensity ($\Delta$) in terms of $\mv$, $\alpha$, $\vmax$, and $\lmax$ as
\be
\label{eq:EvolvedOverdensity}
\Delta = \frac{3\mv}{4\pi\lv^3\rhoc(z)},
\ee
and
\be
\label{eq:EvolvedVirialRadius}
\lv = \frac{\lmax}{(2-\alpha)^2}\frac{\chi_c^2}{(1-\chi_c)^2},\,\,\, {\rm with}\,\,\, \chi_c =\left(\frac{G\mv}{\lmax\vmax^2}\right)^{\frac{1}{2(3-\alpha)}}\left(\frac{2-\alpha}{3-\alpha}\right).
\ee
Using \eqs{EvolvedConcentration}, (\ref{eq:EvolvedOverdensity}), and (\ref{eq:EvolvedVirialRadius}), we can update an evolved \citetalias{Dekel17} subhalo according to the mass $\mv(t)$ from the tidal stripping recipe in \se{stripping} and the evolved structure, $\lmax$ and $\vmax$, from the tidal tracks. 

One can derive equivalent expressions for the \citetalias{Einasto65} profile, linking the \citetalias{Einasto65} concentration, $c_2$, the shape index, $n$, and the overdensity, $\Delta$, to $\vmax$, $\lmax$, and an inner slope, $s(10^{-3}\lv)=2(10^{-3}c)^{1/n}$, which is assumed to be constant.
We omit the derivations here. 


\section{Illustration: evolution of one satellite in a constant potential}
\label{app:illustration}

\begin{figure*}
	\includegraphics[width=\textwidth]{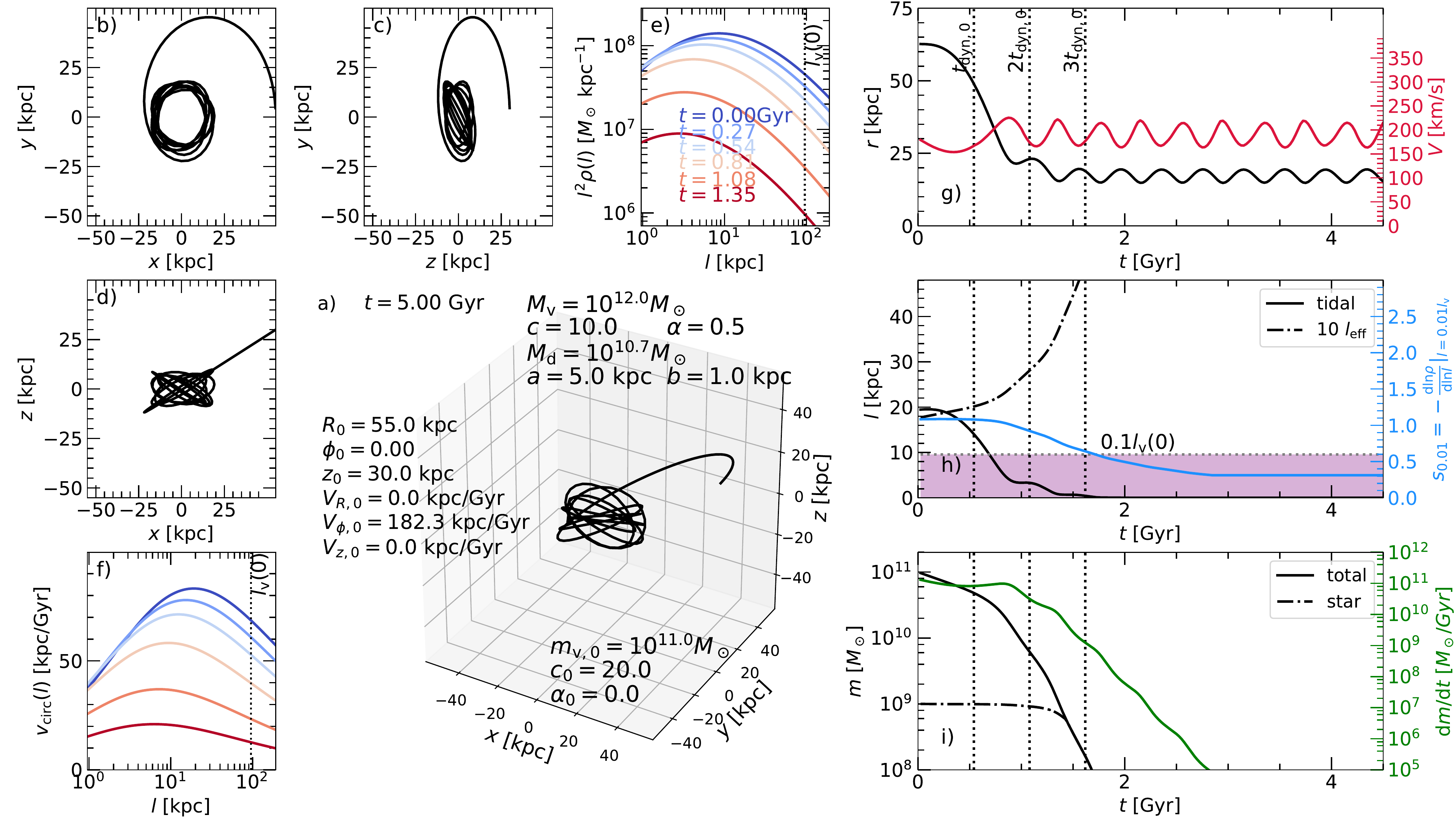}
    \caption{Illustration of satellite evolution in \SatGen: an idealized case where a satellite with initial halo mass of $\mv=10^{11}\Msun$ described by a \citetalias{Dekel17} profile with $c=20$ and $\alpha=0$ (i.e., $c_2 = 11.25$ and $s_{0.01}\approx1.1$) orbits around a central galaxy consisting of a halo of $\Mv=10^{12}\Msun$, $c=10$, and $\alpha=0.5$ (i.e., $c_2 = 10$ and $s_{0.01}=1.22$) and a disc of mass $\Md=10^{10.7}\Msun$ with a scale size of $a=5\kpc$ and a scale height of $b=1\kpc$. 
The satellite is released from $(R,z)=(55,30)$ with a $\boldsymbol{\hat{\phi}}$-direction velocity of approximately the local circular velocity of the host potential and is evolved for 5 Gyr, during which the host potential is fixed (see the text for more details).
Panels (a)-(d) show the orbit in 3D and in the$x-y$, $y-z$, and $x-z$ planes, respectively. 
Panels (e)-(f) show the density profile and circular velocity profile at different epochs, as indicated.
The initial virial radius of the satellite is marked by the vertical dotted line.  
Panels (g)-(i) show the instantaneous values of a few quantities of the satellite as functions of time -- (g) orbital radius and orbital velocity; (h) tidal radius, half-stellar-mass radius, and logarithmic density slope at 0.01$\lv(t)$ (the horizontal dotted line indicates 10\% of the initial virial radius; once the tidal radius drops below this line, the stellar mass loss becomes significant); (i) subhalo mass, stellar mass, and the subhalo mass loss rate.
As a massive satellite, it experiences strong dynamical friction such that its orbit decays by roughly two-thirds in radius in ${\sim}2$ initial, local dynamical times or ${\sim} 1$ Gyr [Panel (e)]. 
It experiences tidal stripping and structural evolution along the way: notably, the maximum circular velocity decreased by roughly one third [Panel (f): the solid lines show the $\vc(l)$ profiles]; the half-stellar-mass radius increased by 50\% [Panel (h), dash-dotted line]; the inner density slope ($s_{0.01}$) decreased from 1.1 to 0.3 [Panel (h), blue line]. 
Afterwards, the disc dominates the dynamics, working to drag the satellite into co-rotation. 
 }
    \label{fig:test_evolve}
\end{figure*}

As an illustration of what has been described in \se{profiles}-\se{tracks}, \fig{test_evolve} presents the evolution of a satellite in a fixed host potential consisting of a \citetalias{Dekel17} halo and a \citetalias{MN75} disc.
The satellite initially has a halo mass of $\mv=10^{11}\Msun$ and is described by a \citetalias{Dekel17} profile with $c=20$ and $\alpha=0$, which corresponds to a conventional concentration of $c_2 = 11.25$ and an inner density slope of $s_{0.01}\approx1.08$. 
It is also initialized with a stellar mass of $\ms=10^9\Msun$ and a half-stellar-mass radius of $\leff=1.6\kpc$. 
The central galaxy has a halo of $\Mv=10^{12}\Msun$, $c=10$, and $\alpha=0.5$ (i.e., $c_2 = 10$ and $s_{0.01}=1.22$), as well as a disk of mass $\Md=10^{10.7}\Msun$ with a scale size of $a=5\kpc$ and a scale height of $b=1\kpc$.
The satellite is released from an off-disc-plane position, $(R,z)=(55,30)$, with an initial velocity that is approximately the local circular velocity in the $\boldsymbol{\hat{\phi}}$ direction. 
All of these are arbitrary choices for illustration purposes.

As can be expected, this massive satellite, with a satellite-to-central mass ratio of ${\sim}0.1$, experiences strong dynamical friction. 
In about two initial, local dynamical times (${\sim}1$ Gyr), its orbital radius decays from the initial ${\sim}60$ kpc to $\la20$ kpc, where it experiences strong tidal stripping, with the instantaneous tidal radius dropping below 10\% of its initial virial radius.
Tidal stripping, heating, and the re-virialization of the satellite is captured by the tidal evolutionary tracks, such that after the ${\sim}1$ Gyr evolution: first, the density profile becomes shallower at $0.01\lv$; second, the maximum circular velocity, $\vmax$, drops from ${\sim}90$ to ${\sim}60$ kpc/Gyr, and the $\vmax$ location, $\lmax$, decreases from 20 kpc to 8 kpc; finally, the half-stellar-mass radius increases from 1.6 kpc to 2.5 kpc. 

Afterwards, the strong mass loss weakens the dynamical friction force and the influence of the disc begins to kick in: the dynamical friction force from the disc works to to drag the satellite into co-rotation, such that after traversing the disk plane several times, the satellite gradually settles into a stable orbit with a radius between 15 and 20 kpc. 


\bsp	
\label{lastpage}
\end{document}